\documentclass[aps]{revtex4}
\usepackage{amssymb,epsf}

\begin{document}

\title{Three dimensional nonlinear magnetic AdS solutions through
topological defects}
\author{S. H. Hendi$^{1,2}$\footnote{
email address: hendi@shirazu.ac.ir}, B. Eslam
Panah$^{1}$\footnote{email address:
behzad$_{-}$eslampanah@yahoo.com}, M.
Momennia$^{1}$\footnote{email address: momennia1988@gmail.com} and
S. Panahiyan$^{1}$\footnote{email address: ziexify@gmail.com}}
\affiliation{$^1$ Physics Department and Biruni Observatory,
College of Sciences, Shiraz University, Shiraz 71454, Iran\\
$^2$ Research Institute for Astronomy and Astrophysics of Maragha (RIAAM),
Maragha, Iran}

\begin{abstract}
Inspired by large applications of topological defects in describing
different phenomena in physics, and considering the importance of three
dimensional solutions in AdS/CFT correspondence, in this paper we obtain
magnetic anti-de Sitter solutions of nonlinear electromagnetic fields. We
take into account three classes of nonlinear electrodynamic models; first
two classes are the well-known Born-Infeld like models including logarithmic
and exponential forms and third class is known as the power Maxwell
invariant nonlinear electrodynamics. We investigate the effects of these
nonlinear sources on three dimensional magnetic solutions. We show that
these asymptotical AdS solutions do not have any curvature singularity and
horizon. We also generalize the static metric to the case of rotating
solutions and find that the value of the electric charge depends on the
rotation parameter. Finally, we consider the quadratic Maxwell invariant as
a correction of Maxwell theory and in other words, we investigate the
effects of nonlinearity as a correction. We study the behavior of the
deficit angle in presence of these theories of nonlinearity and compare them
with each other. We also show that some cases with negative deficit angle
exists which are representing objects with different geometrical structure.
We also show that in case of the static only magnetic field exists whereas
by boosting the metric to rotating one, electric field appear too.
\end{abstract}

\maketitle

\section{Introduction}

Topological defects are playing crucial role in studying different physical
phenomena in context of quantum theory, condensed matter, cosmology and
string theory. It was shown that, in studying liquid crystal, broad range of
phenomena from structural properties to phase transitions are governed by
the existence of these topological defects \cite{Mesaros}. In addition by
employing rules governed by these topological defects some experimental
modifications were done which lead to improvement of physical insight
regarding quantum loops and the quality of obtained materials in different
aspects \cite{Mesaros,Tkalec}. These studies are acceptable in context of
condensed matter with ordered media \cite{Mermin}. Another application of
these defects is in studying magnetism and nanomagnetism which was done in
literature \cite{Braun}. In addition, it is notable that this mathematical
tool was also employed in studying vortices in superfluid \cite{Aad} and
Bose-Einstein condensate \cite{Simula,Sabbatini}. Also, in studying the
phase transition and critical behavior of Bose gas \cite{Navon}, it was used
widely. Furthermore, it is worthwhile to mention that these topological
defects are essential tools in order to study superconductors and their
phase transitions \cite{Kierfeld}.

In general a topological defect or topological solution is formed
because of symmetry breakdown. Due to different symmetry
breakdowns and their dimensionality, the topological defects could
be interpreted as different types of known defects. In cosmology,
the well-known topological defects are cosmic string \cite{CS},
domain walls \cite{DW}, monopoles \cite{Mon}, textures \cite{Tex}
and (higher dimensional) branes (for more details see
\cite{Brandenberger}). Certain theories of GUT predict that these
cosmological defects are formed during phase transition of early
Universe \cite{Kibble} and they can be used to describe the large
scale structure of the Universe. As for cosmic string,
asymptotically AdS spacetimes generated by static and spinning
magnetic sources in the Einstein-Maxwell gravity have been
investigated in \cite{OJCDias}. The properties and interaction of
the superconducting cosmic strings with astrophysical magnetic
fields have been studied in \cite{Witten1985}. Also,
superconducting cosmic strings have been investigated in dilaton
gravity \cite{CNFerreira}, and in Brans-Dicke theory \cite{AASen}.
These solutions are not black holes, and represent spacetimes with
conic singularities. From cosmological point of view, the
properties of the magnetic (cosmic) string have been studied in
many literatures (for e.g., see \cite{Brand1992}). Properties of
the QCD static strings and applications of magnetic strings in
quantum theories have been explored in \cite{QCD} and
\cite{Quantum}, respectively. The main reason to consider the
magnetic string solutions is that they may be interpreted as
cosmic strings.
As for the domain walls, it has been studied in many literature \cite%
{Skenderis} and for textures some studies have been done
\cite{Phillips}. In gravitational point of view, these topological
defects are characterized by their masses, rotation parameters and
deficit angles, while from cosmological aspect, these topological
defects, as predicted, have no gravitational potential and the
only property that makes them visible or detectable is their
deficit angle which is acting as a cosmological lens
\cite{Sazhin}. The projection of photon on the surface of these
topological defects is modified due to existence of deficit angle.
Therefore, in case of these cosmological objects, the property
known as deficit angle is so important.

Motivated by the above statements, in this paper we study three
dimensional AdS magnetic solutions in context of topological
defects in the presence of different nonlinear theories of
electrodynamics. As we will see, these solutions do not have any
singularity and horizon. Therefore, these solutions are not
interpreted as black holes, but spacetimes with a conical
singularity. In other words, our solutions are the nonlinear
counterparts of the static Einstein-Maxwell-AdS solutions found in
Ref. \cite{StaticSol} and of the rotating solutions in Ref.
\cite{Dias}.

Most of physical systems have a nonlinear behavior in nature, so, the
nonlinear field theories are interesting in physics. The basic motivation
for studying the nonlinear electrodynamics (NED) comes from the fact that
these theories are generalizations of the Maxwell field and in the special
case (weak nonlinearity)\ they can reduce to the linear Maxwell theory.
Other motivations of considering NED are limitations of the Maxwell theory
\cite{Maxwell}, description of the self-interaction of virtual
electron-positron pairs \cite{Heisenberg} and the radiation propagation
inside specific materials \cite{materials}. Besides, NED improves the basic
concept of gravitational redshift and its dependency of any background
magnetic field as compared to the well-established method introduced by
standard general relativity. In addition, it was recently shown that NED
objects can remove both of the big bang and black hole singularities \cite%
{singularities}. Moreover, from astrophysical point of view, one finds the
effects of NED become indeed quite important in superstrong magnetized
compact objects, such as pulsars, and particular neutron stars \cite%
{neutronstars}.

It is well-known that the electric field of a point-like charge has a
singularity in its location (origin). In order to remove this divergency,
Born and Infeld introduced a NED which is known as Born-Infeld nonlinear
electrodynamics (BI NED) theory \cite{BI}. After that Soleng and Hendi
introduced two different types of BI type NED in \cite{Soleng}\ and \cite%
{HendisTheory}, respectively, which can also remove the electric field
divergency of point-like charges near the origin. Soleng's Lagrangian has a
logarithmic form and, like BI NED, removes divergences in the electric
field, while Hendi's Lagrangian has an exponential form and does not cancel
the origin divergency of the electric field but its singularity is much
weaker than Einstein-Maxwell theory. Another example of the nonlinear
electromagnetic field is power Maxwell invariant (PMI NED)\ field \cite{PMI}%
. In general BI-types of NED theories have interesting properties that make
them different comparing to other theories of nonlinearity \cite{Boillat}.
In addition in context of heterotic string theory in low energy limit, the
Lagrangian of these kinds of nonlinear theories may rise which gives another
strong motivation for considering these theories.

On the other hand, one of the main reasons to consider the $(2+1)$%
-dimensional solutions becomes from the fact that these solutions help us to
find a profound insight in the physics of $(2+1)$-dimensional objects and
also, play an important role to improve our understanding of gravitational
interaction in low-dimensional manifolds. Due to these facts, investigation
of $(2+1)$-dimensional spacetimes is important. Three dimensional solutions
of black holes and magnetic solutions have been studied by many authors \cite%
{Three,Hendi2014,StaticSol,Dias}. This tremendous interest in
these solutions is due to fact that three dimensional solutions
contribute fundamentally to conceptual issues of astrophysical
subjects such as black holes thermodynamics \cite{Carlip}. In
addition, in context of quantum gravity and string theory, also
due to AdS/CFT correspondence, these theories play an undeniable
important role in describing different phenomena specially in low
dimensional spacetime \cite{Witten}. Recently, some theories have
been proposed for obtaining magnetic solutions in three dimensions
and magnetic monopole \cite{Mazharimousavi}.

The outline of our paper is as follows. In the next section, we give a brief
review of the basic field equations of Einstein gravity in the presence of
cosmological constant and nonlinear electrodynamics. In Sec. \ref{Sol}, we
consider the $(2+1)$-dimensional horizonless metric and obtain magnetic
solutions for various sources and investigate its geometric properties.
Also, we apply the rotation boost to the static metric and obtain the
conserved quantities of rotating spacetime. Next, we consider nonlinearity
as a correction to Maxwell theory and study the magnetic solutions for this
case. Geometrical properties of solution will be studied. We finish our
paper with some concluding remarks.

\section{Basic Field Equations}

\label{FE}

The $(2+1)$-dimensional action in Einstein gravity with negative
cosmological constant ($\Lambda $) that coupled to nonlinear electrodynamics
is given by%
\begin{equation}
\mathcal{I}_{G}=-\frac{1}{16\pi }\int_{\mathcal{M}}d^{3}x\sqrt{-g}\left[
\mathcal{R}-2\Lambda +\mathcal{L}_{NL}(\mathcal{F})\right] -\frac{1}{8\pi }%
\int_{\partial \mathcal{M}}d^{2}x\sqrt{-\gamma }\Theta \left( \gamma \right)
,  \label{Action}
\end{equation}%
where $\mathcal{R}$\ is the scalar curvature and $\mathcal{L}_{NL}(\mathcal{F%
})$\ is an arbitrary Lagrangian of nonlinear electrodynamics. The last term
in the present equation is the Gibbons-Hawking surface term in which we must
add it to the action for a well-defined variational principle \cite{Myers,
Davis}. In this term $\gamma $\ and $\Theta $\ are, respectively, the trace
of the induced\ metric $\gamma _{ij}$\ and the extrinsic curvature $\Theta
_{ij}$\ on the boundary $\partial \mathcal{M}$. Varying the action (\ref%
{Action}) with respect to the gravitational field $g_{\mu \nu }$\ and the
gauge potential $A_{\mu }$, the field equations are obtained as%
\begin{equation}
R_{\mu \nu }-\frac{1}{2}g_{\mu \nu }\left( \mathcal{R}-2\Lambda \right)
=T_{\mu \nu },  \label{Field equation}
\end{equation}%
\begin{equation}
\partial _{\mu }\left( \sqrt{-g}\mathcal{L}_{\mathcal{F}}F^{\mu \nu }\right)
=0,  \label{Maxwell equation}
\end{equation}%
where $\mathcal{L}_{\mathcal{F}}=d\mathcal{L}_{NL}(\mathcal{F})/d\mathcal{F}$
and $\mathcal{F}=F_{\mu \nu }F^{\mu \nu }$\ is the Maxwell invariant where $%
F_{\mu \nu }$\ is the electromagnetic tensor field. In the presence of
nonlinear electromagnetic field, the energy-momentum tensor of Eq. (\ref%
{Field equation}) is%
\begin{equation}
T_{\mu \nu }=\frac{1}{2}g_{\mu \nu }\mathcal{L}_{NL}(\mathcal{F})-2\mathcal{L%
}_{\mathcal{F}}F_{\mu \lambda }F_{\nu }^{\lambda }.  \label{Energy momentum}
\end{equation}

In general the action $\mathcal{I}_{G}$ diverges when evaluated on the
solutions, as the Hamiltonian and other associated conserved quantities.
Rather than eliminating these divergences by incorporating reference term, a
counterterm action $\mathcal{I}_{ct}$ is added to the action which is
functional of the boundary curvature invariants. For asymptotically AdS
solutions, one can instead deal with these divergences via the counterterm
method inspired by AdS/CFT correspondence \cite{Mald}. We assume that the
suitable counterterm is
\begin{equation}
\mathcal{I}_{ct}=-\frac{1}{8\pi }\int_{\partial \mathcal{M}}d^{2}x\sqrt{%
-\gamma }\mathcal{L}_{ct},
\end{equation}
where $\mathcal{L}_{ct}$ is the counterterm Lagrangian and by use of the
suitable Lagrangian we will be able to compute the finite conserved
quantities. Therefore, the total finite action, $\mathcal{I}$, can be
written as
\begin{equation}
\mathcal{I}=\mathcal{I}_{G}+\mathcal{I}_{ct}.  \label{Total action}
\end{equation}

Having the total finite action, one can use Brown and York definition \cite%
{Brown} to construct a divergence free stress-energy tensor as
\begin{equation}
T^{\mu \nu }=\frac{1}{8\pi }\left( \Theta ^{\mu \nu }-\Theta \gamma ^{\mu
\nu }+2\frac{\delta \mathcal{L}_{ct}}{\delta \gamma _{\mu \nu }}\right) .
\label{stress-energy tensor}
\end{equation}

To compute the conserved charges of a rotating spacetime, we choose a
spacelike surface $\mathcal{B}$ in $\partial \mathcal{M}$ with metric $%
\sigma $ , and write the boundary metric in ADM (Arnowitt-Deser-Misner) form
\begin{equation}
\gamma _{\mu \nu }dx^{\mu }dx^{\nu }=-N^{2}dt^{2}+\sigma \left( d\varphi
+Vdt\right) ^{2},  \label{ADM form}
\end{equation}
where the coordinates $\varphi$ is the angular variables parameterizing the
hypersurface of constant radial coordinate around the origin, and $N$ and $V$
are the lapse and shift functions, respectively. Considering a Killing
vector field $\xi$ on the boundary, then the quasilocal conserved quantities
associated with the stress energy momentum tensor of Eq. (\ref{stress-energy
tensor}) can be written as
\begin{equation}
Q\left( \xi \right) =\int_{\mathcal{B}}d\varphi \sqrt{\sigma }T_{\mu \nu
}n^{\mu }\xi ^{\nu },  \label{CQ}
\end{equation}%
where $n^{\mu }$ is the timelike unit normal vector to the boundary $%
\mathcal{B}$. For boundaries with timelike ($\xi =\partial /\partial t$) and
rotational ($\varsigma =\partial /\partial \varphi $) Killing vector fields,
one can obtain associated conserved quantities in the following form
\begin{equation}
M=\int_{\mathcal{B}}d\varphi \sqrt{\sigma }T_{\mu \nu }n^{\mu }\xi ^{\nu },
\label{M}
\end{equation}
\begin{equation}
J=\int_{\mathcal{B}}d\varphi \sqrt{\sigma }T_{\mu \nu }n^{\mu }\varsigma
^{\nu },  \label{J}
\end{equation}
provided the surface $\mathcal{B}$ contains the orbits of $\varsigma$. These
quantities are the mass and angular momentum of the system enclosed by the
boundary $\mathcal{B}$, respectively.

\section{Magnetic Solutions with Nonlinear Sources}

\label{Sol}

In this section we want to obtain the three dimensional solutions of Eqs. (%
\ref{Field equation})-(\ref{Energy momentum}) with considering different
electrodynamic models. We consider the following ansatz for the metric
\begin{equation}
ds^{2}=-\frac{\rho ^{2}}{l^{2}}dt^{2}+\frac{d\rho ^{2}}{g(\rho )}
+l^{2}g(\rho )d\varphi ^{2},  \label{metric}
\end{equation}
where $g(\rho)$ is an arbitrary function of radial coordinate, $\rho$, and
should be determined and $l$ is a scale length factor which is related to $%
\Lambda $. The angular coordinate $\varphi$ is dimensionless and ranges in $%
0 \leq \varphi <2 \pi$. The motivation for this curious choice for the
metric gauge [$g_{tt}\varpropto -\rho ^{2}$ and $\left( g_{\rho \rho
}\right) ^{-1}\varpropto g_{\varphi \varphi }$] instead of the usual
Schwarzschild like [$\left( g_{\rho \rho }\right) ^{-1}\varpropto g_{tt}$
and $g_{\varphi \varphi }\varpropto \rho ^{2}$] comes from the fact that we
are looking for magnetic solutions. It is easy to show that, using a
suitable transformation, the metric (\ref{metric}) can map to $3$%
-dimensional Schwarzschild like spacetime locally, but not globally.

It is well-known that the electric field is associated with the time
component of the vector potential, $A_{t}$, while the magnetic field is
associated with the angular component $A_{\varphi }$. Since we want to
investigate the magnetic solutions, so we assume the vector potential as
\begin{equation}
A_{\mu }=\left(\int F_{\varphi \rho} d\rho \right)\delta _{\mu }^{\varphi }.
\label{Gauge potential}
\end{equation}
Now we continue our paper for obtaining the magnetic solutions in the
Einstein gravity and in presence of various models of NED.

\subsection{$\emph{Static~Solutions}$}

\subsubsection{Class I: PMI NED model}

In this case, we want to obtain the solutions in presence of PMI NED and
investigate the properties of the solutions. Therefore, we consider the PMI
Lagrangian with the following form%
\begin{equation}
\mathcal{L}_{PMI}(\mathcal{F})=(-\kappa \mathcal{F})^{s},  \label{PMI}
\end{equation}%
where $\kappa $ and $s$ are coupling and arbitrary constants, respectively.
It is straightforward to show that for $s=1$, the PMI Lagrangian (\ref{PMI})
reduces to the standard Maxwell Lagrangian ($\mathcal{L}_{Maxwell}(\mathcal{F%
})=-\kappa \mathcal{F}$). Since the Maxwell invariant is negative, hereafter
we set $\kappa =1$, without loss of generality. Using the nonlinear Maxwell
equation (\ref{Maxwell equation}) and the Lagrangian of PMI (\ref{PMI}) with
the metric (\ref{metric}), one can obtain
\begin{equation}
F_{\varphi \rho }+K(\rho )=0,  \label{Fpr}
\end{equation}%
where%
\begin{equation}
K(\rho )=(2s-1)\rho F_{\varphi \rho }^{\prime },
\end{equation}%
where the "prime" denotes differentiation with respect to $\rho $. Eq. (\ref%
{Fpr}) has the following solutions
\begin{equation}
F_{\varphi \rho }=\frac{q}{\rho ^{1/(2s-1)}},  \label{PMI field}
\end{equation}%
where $q$ is an integration constant. In order to have physical asymptotical
behavior, we restrict ourselves to $s>1/2.$ To find the metric function $%
g(\rho )$, one may insert Eqs. (\ref{PMI field}) and (\ref{metric}) in the
field equation (\ref{Field equation}). After some calculations, one can show
that
\begin{equation}
\left\{
\begin{array}{cc}
g^{\prime }(\rho )+2\Lambda \rho -(2s-1)\left( \frac{2q^{2}}{l^{2}\rho
^{1/s(2s-1)}}\right) ^{s}=0, & \rho \rho \ (\varphi \varphi )\ \text{%
component} \\
g^{\prime \prime }(\rho )+2\Lambda +\left( \frac{2q^{2}}{l^{2}\rho
^{2/(2s-1)}}\right) ^{s}=0, & tt\ \text{component}%
\end{array}%
\right. ,
\end{equation}

It is straightforward to show that these equations have the following
solutions
\begin{equation}
g(\rho )=m-\Lambda \rho ^{2}+\left\{
\begin{array}{cc}
\frac{2q^{2}\ln \left( \frac{\rho }{l}\right) }{l^{2}}, & s=1 \\
\frac{2^{s-1}(2s-1)^{2}}{\left( s-1\right) }\left( \frac{q}{l}\right)
^{2s}\rho ^{2(s-1)/(2s-1)}, & otherwise%
\end{array}%
\right. ,  \label{PMI metric}
\end{equation}%
where $m$ is the integration constant which is related to the mass parameter.

\subsubsection{Class II: Exponential form of NED (ENED)}

Here, we consider ENED Lagrangian as
\begin{equation}
\mathcal{L}_{\exp }(\mathcal{F})=\beta ^{2}\left( \exp \left( -\frac{%
\mathcal{F}}{\beta ^{2}}\right) -1\right) ,  \label{ENED}
\end{equation}%
where $\beta $\ is the ENED parameter and in the limit $\beta
\longrightarrow \infty $, $\mathcal{L}_{\exp }(\mathcal{F})$ reduces to the
standard Maxwell form $\mathcal{L}_{Maxwell}(\mathcal{F})=-\mathcal{F}$.
Inserting the Lagrangian of ENED (\ref{ENED}) in the nonlinear Maxwell
equation (\ref{Maxwell equation}) and using the metric (\ref{metric}), one
can obtain
\begin{equation}
\left[ 1-\left( \frac{2F_{\varphi \rho }}{l\beta }\right) ^{2}\right]
F_{\varphi \rho }^{\prime }+\frac{F_{\varphi \rho }}{\rho }=0.
\end{equation}

This equation has the following solution
\begin{equation}
F_{\varphi \rho }=\frac{l\beta }{2}\sqrt{-L_{W}},  \label{ENEF
field}
\end{equation}%
where $L_{W}=LambertW\left( -4q^{2}/l^{2}\beta ^{2}\rho ^{2}\right) $ and
the parameter $q$ is an integration constant. It is worthwhile to note that
in order to have a real electromagnetic field, we should consider $\rho $
with the following limitation
\begin{equation}
\rho >\rho _{0}=\frac{2q}{l\beta }\exp \left( \frac{1}{2}\right) .
\end{equation}

Now, we want to obtain the function of $f\left( \rho \right) $. For this
purpose, we can take into account (\ref{metric}) and (\ref{ENEF field}) in
the gravitational field equation (\ref{Field equation}) to obtain its
nonzero components as
\begin{equation}
\left\{
\begin{array}{cc}
g^{\prime }\left( \rho \right) +\rho \left( 2\Lambda +\beta ^{2}-\frac{%
2\beta q}{l\rho }\left[ \left( -L_{W}\right) ^{-1/2}-\left( -L_{W}\right)
^{1/2}\right] \right) =0, & \rho \rho \ (\varphi \varphi )\ \text{component}
\\
g^{\prime \prime }\left( \rho \right) +\left( 2\Lambda +\beta ^{2}-\frac{%
2\beta q}{l\rho }\left( -L_{W}\right) ^{-1/2}\right) =0, & tt\ \text{%
component}%
\end{array}%
\right. ,
\end{equation}

After some calculations, one can show that these equations have the
following solution
\begin{equation}
g\left( \rho \right) =m-\Lambda \rho ^{2}-\frac{\beta ^{2}\rho ^{2}}{2}+\int
\frac{2\beta q}{l}\left( \sqrt{-L_{W}}+\frac{1}{\sqrt{-L_{W}}}\right) d\rho .
\label{ENEF
metric}
\end{equation}

\subsubsection{Class III: Logarithmic form of NED (LNED)}

Now, we want to consider the LNED Lagrangian with the following form
\begin{equation}
\mathcal{L}_{\log }(\mathcal{F})=-8\beta ^{2}\ln \left( 1+\frac{\mathcal{F}}{%
8\beta ^{2}}\right) ,  \label{LNED}
\end{equation}%
where $\beta $ is the LNED parameter and for weak nonlinearity limit $\beta
\longrightarrow \infty $, $\mathcal{L}_{\log }(\mathcal{F})$ reduces to the
standard Maxwell form $\mathcal{L}_{Maxwell}(\mathcal{F})=-\mathcal{F} $.
Using the LNED Lagrangian (\ref{LNED}) and the nonlinear Maxwell equation (%
\ref{Maxwell equation}) with the metric (\ref{metric}), leads to
\begin{equation}
\left[ 1-\left( \frac{F_{\varphi \rho }}{l\beta }\right) ^{2}\right]
F_{\varphi \rho }^{\prime }+\left[ 1+\left( \frac{F_{\varphi \rho }}{l\beta }%
\right) ^{2}\right] \frac{F_{\varphi \rho }}{\rho }=0,
\end{equation}%
with the following solution
\begin{equation}
F_{\varphi \rho }=\frac{\rho l^{2}\beta ^{2}}{2q}\left( 1-\Gamma \right) ,
\label{LNEFfield}
\end{equation}%
where $\Gamma =\sqrt{1-\left( 2q/\rho l\beta \right) ^{2}}$. In order to
have a real electromagnetic field, we should consider $\rho $ with the
following restriction
\begin{equation}
\rho >\rho _{0}=\frac{2q}{l\beta }.
\end{equation}

Here, we want to obtain the solutions of Eqs. (\ref{Field equation}) and (%
\ref{Maxwell equation}). Considering Eqs. (\ref{metric}) and (\ref{LNED}),
one can obtain the nonzero components as
\begin{equation}
\left\{
\begin{array}{cc}
g^{\prime \prime }\left( \rho \right) +2\Lambda +8\Pi \beta ^{2}=0, & tt\
\text{component} \\
\left[ g^{\prime }\left( \rho \right) +2\Lambda \rho +8\rho \beta ^{2}\left(
\Pi -2\right) \right] \left( \Gamma -1\right) +\frac{32q^{2}}{\rho l^{2}}=0,
& \rho \rho \ (\varphi \varphi )\ \text{component}%
\end{array}%
\right. ,
\end{equation}%
where%
\begin{equation}
\Pi =\ln \left( \frac{l^{2}\rho ^{2}\beta ^{2}(1-\Gamma )}{2q^{2}}\right) .
\end{equation}

After some calculations, one can determine the metric function $g\left( \rho
\right)$ as
\begin{equation}
g(\rho )=m-\Lambda \rho ^{2}+\frac{8q^{2}}{l^{2}}\ln \left[ l\rho \beta
\left( 1+\Gamma \right) \right] -2\rho ^{2}\beta ^{2}\left[ 3\left( 1-\Gamma
\right) +2\Pi \right] .  \label{LNEF metric}
\end{equation}

\subsection{Energy Condition}

In order to find a physical solutions, we examine the energy conditions for
these nonlinear models. It is usual to consider the orthonormal
contravariant basis vectors and calculate the three dimensional energy
momentum tensor as $T^{\mu \nu }=diag(\mu,p_{r},p_{t})$. The physical
concepts of $\mu$, $p_{r}$ and $p_{t}$ are, respectively, the energy
density, the radial pressure and the tangential pressure. Having the energy
momentum tensor at hand, we are in a position to discuss energy conditions.
We use the following known constraints in three dimensions

\begin{center}
\begin{tabular}{cc}
\hline\hline
$%
\begin{array}{c}
p_{r}+\mu \geq 0 \\
p_{t}+\mu \geq 0%
\end{array}%
,$ & $\text{for null energy condition (NEC)}$ \vspace{0.2cm} \\ \hline
$%
\begin{array}{c}
\mu \geq 0 \\
p_{r}+\mu \geq 0 \\
p_{t}+\mu \geq 0%
\end{array}%
,$ & $\text{for weak energy condition (WEC)}$ \vspace{0.2cm} \\ \hline
$%
\begin{array}{c}
\mu \geq 0 \\
-\mu \leq p_{r}\leq \mu \\
-\mu \leq p_{t}\leq \mu%
\end{array}%
,$ & $\text{for dominant energy condition (DEC)}$ \vspace{0.2cm} \\ \hline
$%
\begin{array}{c}
p_{r}+\mu \geq 0 \\
p_{t}+\mu \geq 0 \\
\mu +p_{r}+p_{t}\geq 0%
\end{array}%
,$ & $\text{for strong energy condition (SEC)}$ \\ \hline\hline
\end{tabular}
\\[0pt]
\vspace{0.1cm} Table ($1$): Energy conditions criteria \vspace{0.5cm}
\end{center}

In order to simplify the mathematics and physical interpretations, we use
the following orthonormal contravariant (hatted) basis vectors for diagonal
static metric (\ref{metric})
\begin{equation}
\mathbf{e}_{\widehat{t}}=\frac{l}{\rho }\frac{\partial }{\partial t},\text{
\ \ }\mathbf{e}_{\widehat{\rho }}=\sqrt{g}\frac{\partial }{\partial \rho },\
\ \mathbf{e}_{\widehat{\phi }}=\frac{1}{l\sqrt{g}}\frac{\partial }{\partial
\phi }.  \label{basis}
\end{equation}%
It is a matter of straightforward calculations to show that the nonzero
components of stress-energy tensor for the mentioned models are
\[
T^{_{\widehat{t}\widehat{t}}}=\left\{
\begin{array}{cc}
\frac{1}{2}\left( \frac{2F_{\phi \rho }^{2}}{l^{2}}\right) ^{s}, & PMI \\
\frac{\beta ^{2}}{2}\left[ 1-\exp \left( \frac{-2F_{\phi \rho }^{2}}{%
l^{2}\beta ^{2}}\right) \right] , & ENED \\
4\beta ^{2}\ln \left[ 1+\left( \frac{F_{\phi \rho }}{2l\beta }\right) ^{2}%
\right] , & LNED%
\end{array}%
\right.
\]%
\[
T^{_{\widehat{\rho }\widehat{\rho }}}=T^{_{\widehat{\phi }\widehat{\phi }%
}}=\left\{
\begin{array}{cc}
\frac{2s-1}{2}\left( \frac{2F_{\phi \rho }^{2}}{l^{2}}\right) ^{s}, & PMI \\
-\frac{\beta ^{2}}{2}\left[ 1-\left( 1+\frac{4F_{\phi \rho }^{2}}{l^{2}\beta
^{2}}\right) \exp \left( \frac{-2F_{\phi \rho }^{2}}{l^{2}\beta ^{2}}\right) %
\right] , & ENED \\
-4\beta ^{2}\left( \ln \left[ 1+\left( \frac{F_{\phi \rho }}{2l\beta }%
\right) ^{2}\right] -\frac{F_{\phi r}^{2}}{2l^{2}\beta ^{2}\left[ 1+\left(
\frac{F_{\phi \rho }}{2l\beta }\right) ^{2}\right] }\right) , & LNED%
\end{array}%
\right. .
\]%
After some calculations, one finds that NEC, WEC and SEC are satisfied for
all models, simultaneously. In addition, it is easy to show that DEC is
satisfy for both ENED and LNED branches, while for PMI case it can be
satisfied for $\frac{1}{2}<s\leq 1$.

In order to investigate the effect of the nonlinearity of the models on the
energy density of the spacetime, we plot the $T^{_{\widehat{t}\widehat{t}}}$%
\ versus $\rho >\rho _{0}$\ for different values of nonlinearity parameter ($%
s$ or $\beta $).
%%%%%%%%%%%%%%%%%%%%%%%%%%%%%%%%%%%%%%%%%%%%%%%%%%%%%%%%%%%%%%%%%%%%
\begin{figure}[tbp]
$%
\begin{array}{ccc}
\epsfxsize=7cm \epsffile{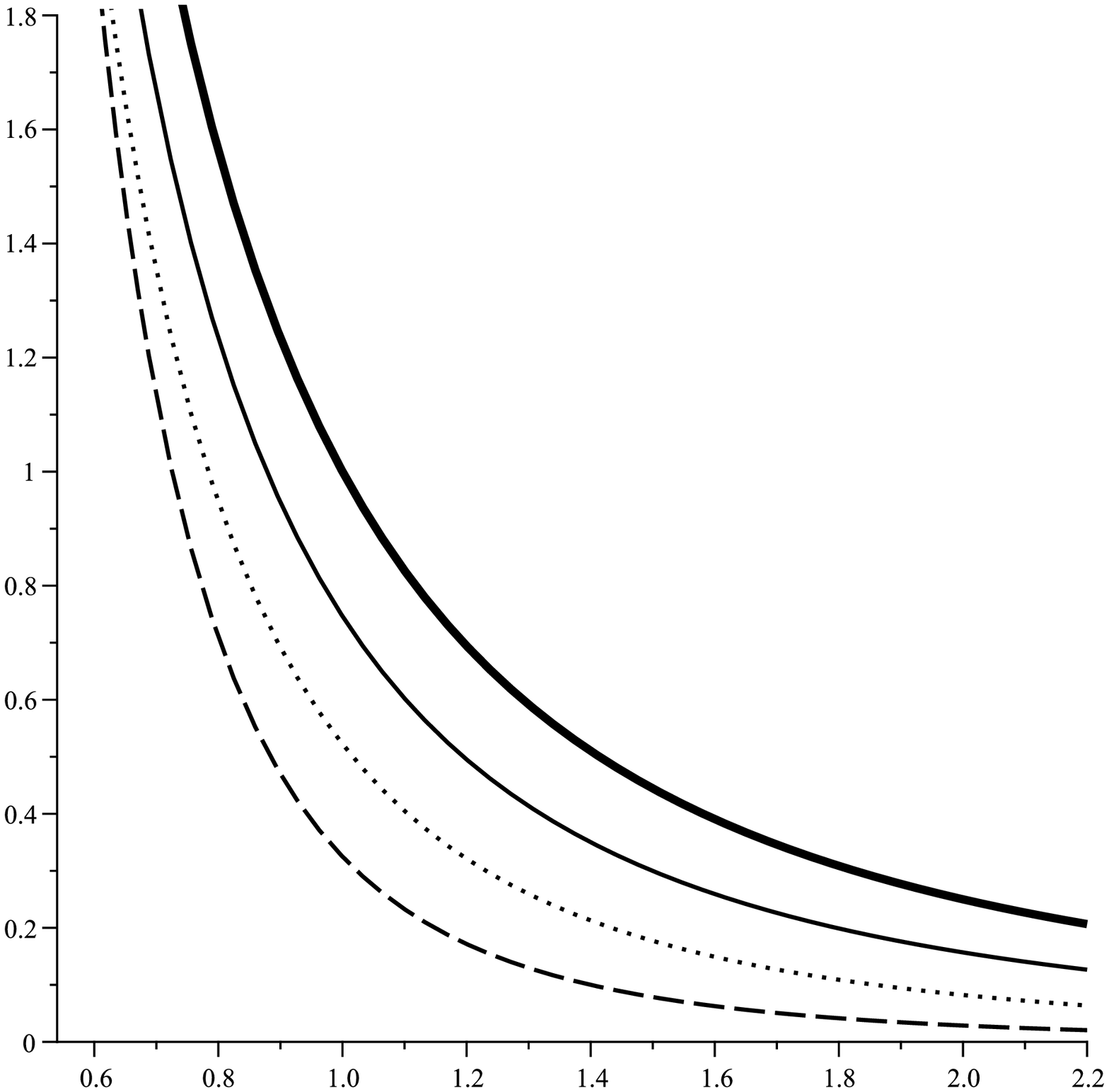} & \epsfxsize=7cm %
\epsffile{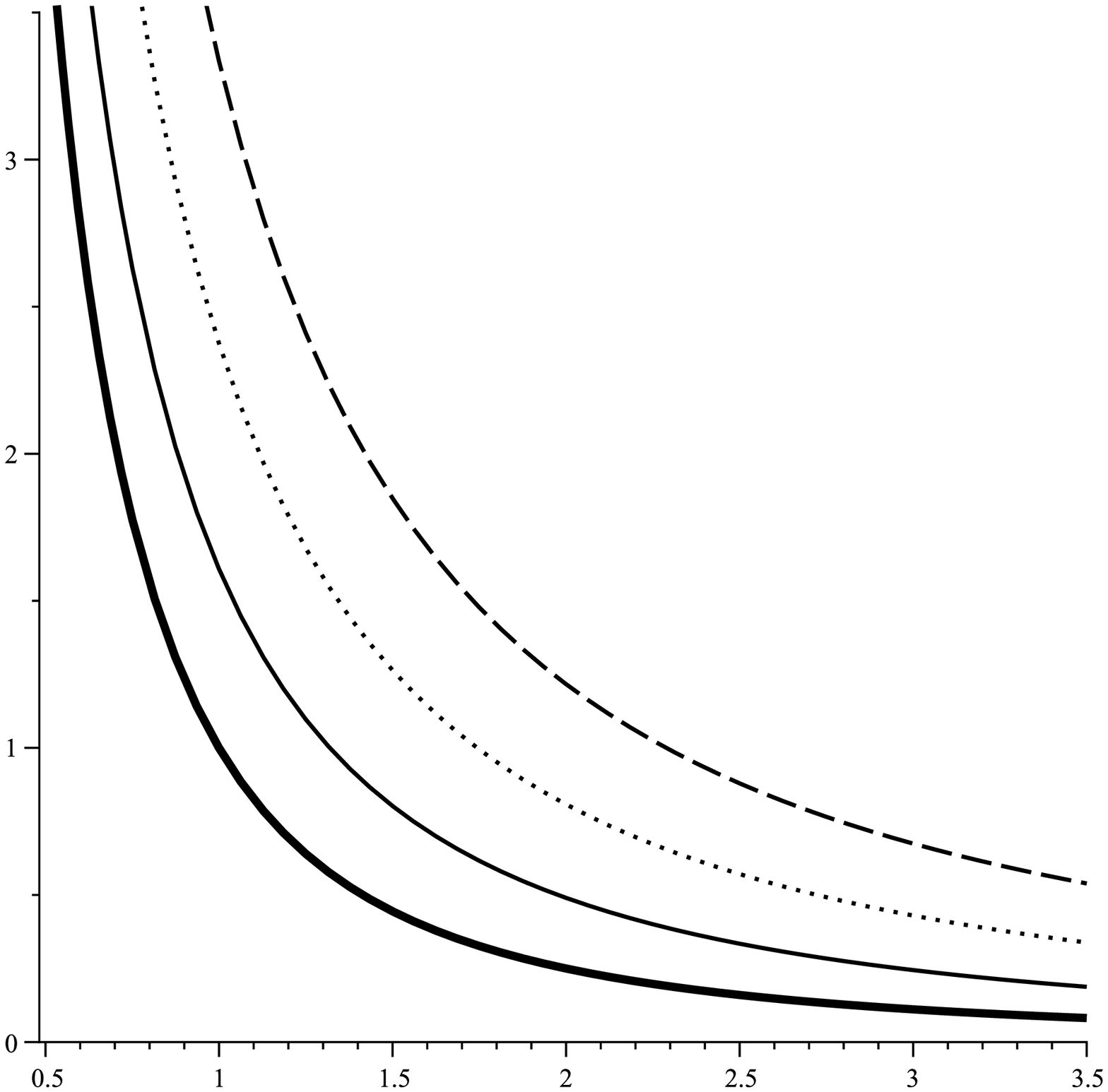} &
\end{array}
$%
\caption{\textbf{\emph{Maxwell and PMI solutions:}} $T^{_{\widehat{\protect%
\rho }\widehat{\protect\rho }}}$ versus $\protect\rho$ for $l=1$ and $q=1$.
\newline
\textbf{Left diagram:} $s=0.7$ (dashed line), $s=0.8$ (doted line), $s=0.9$
(continuous line) and $s=1$ (bold line). \newline
\textbf{Right diagram:} $s=1$ (bold line), $s=1.2$ (continuous line), $s=1.4$
(doted line) and $s=1.6$ (dashed line).}
\label{Fig1}
\end{figure}
%%%%%%%%%%%%%%%%%%%%%%%%%%%%%%%%%%%%%%%%%%%%%%%%%%%%%%%%%%%%%%%%%%%%
%%%%%%%%%%%%%%%%%%%%%%%%%%%%%%%%%%%%%%%%%%%%%%%%%%%%%%%%%%%%%%%%%%%%
\begin{figure}[tbp]
$%
\begin{array}{ccc}
\epsfxsize=7cm \epsffile{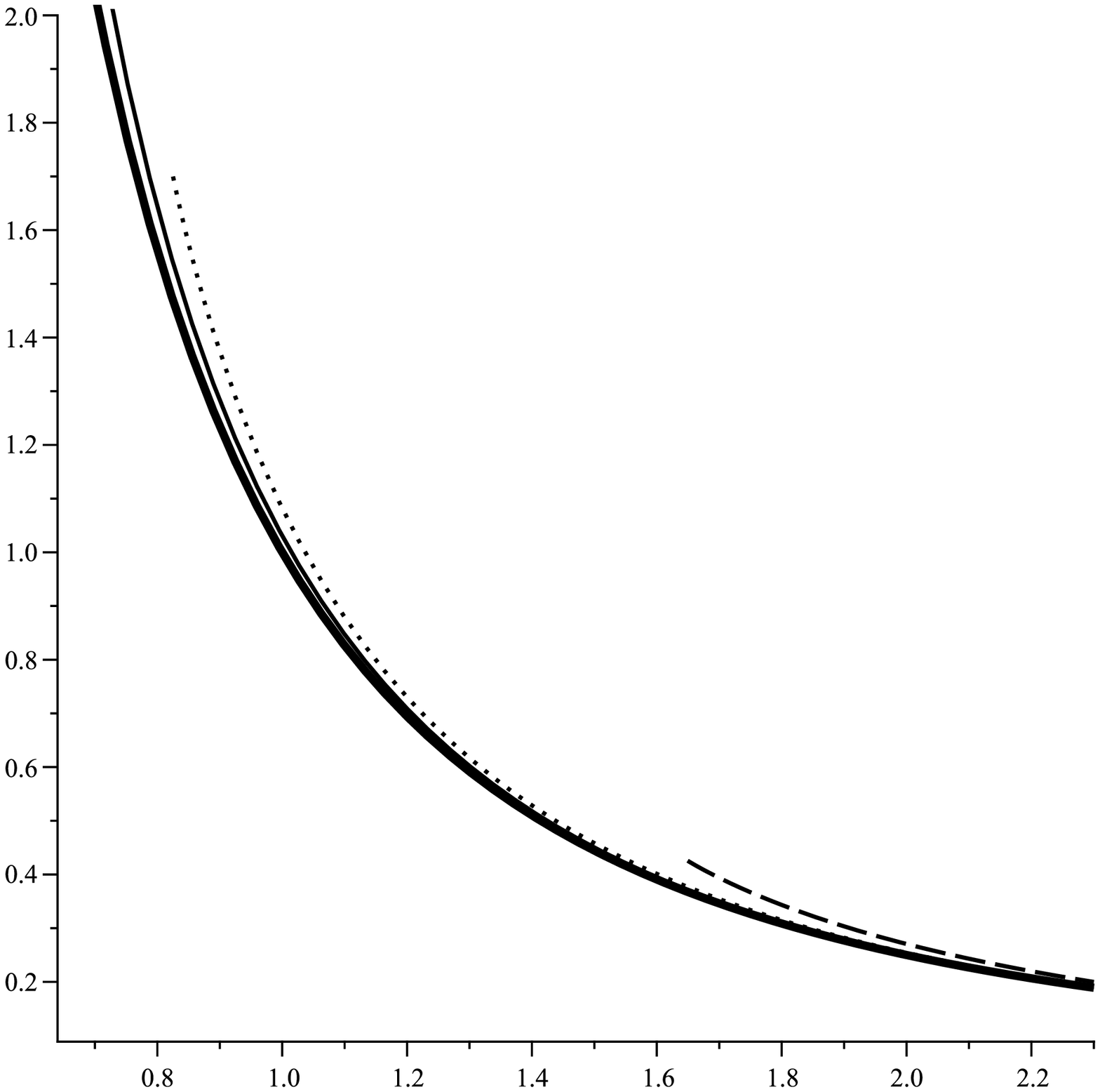} & \epsfxsize=7cm \epsffile{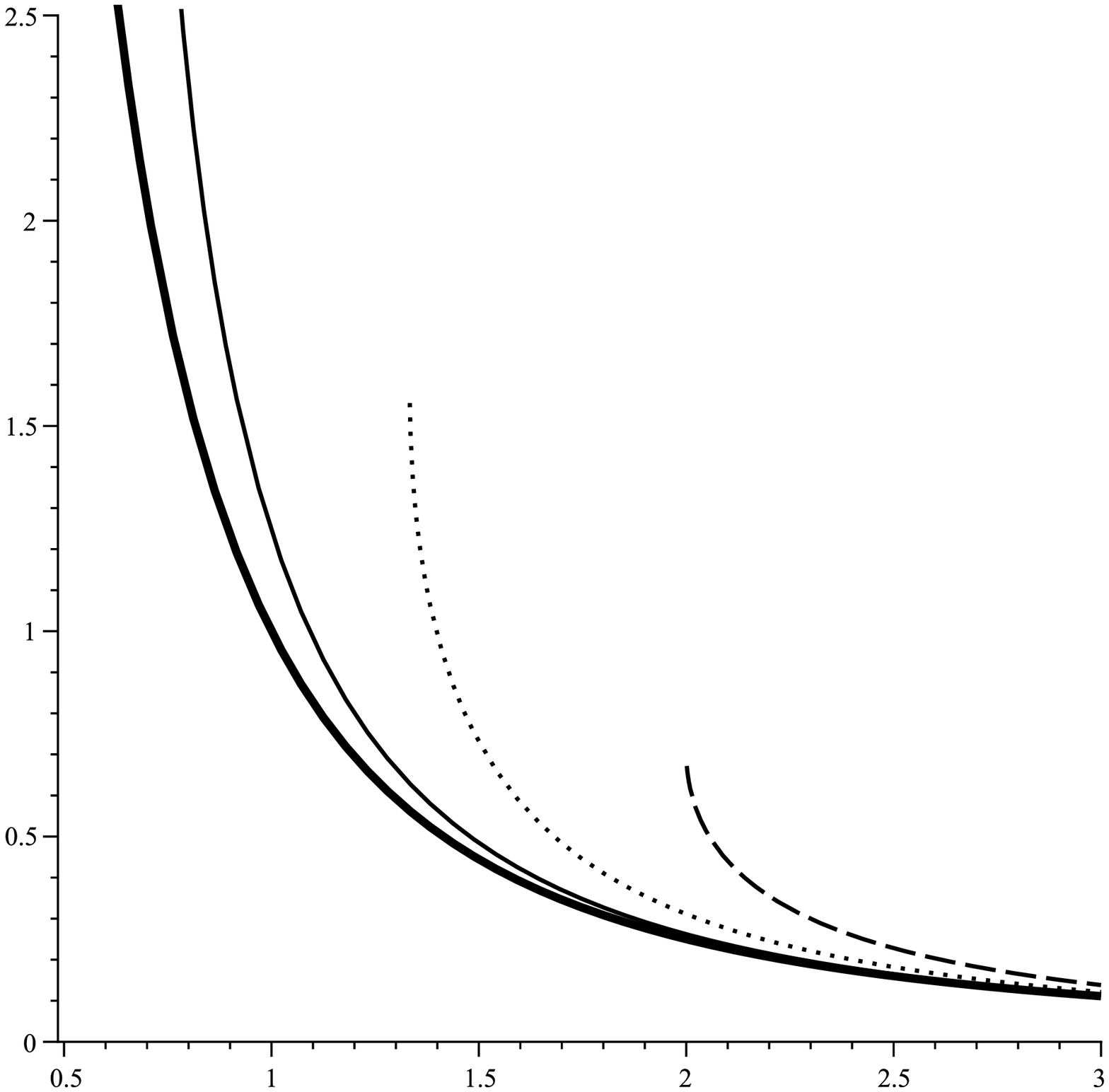}
&
\end{array}
$%
\caption{\textbf{\emph{ENED and LNED solutions:}} $T^{_{\widehat{\protect%
\rho }\widehat{\protect\rho }}}$ versus $\protect\rho$ for $l=1$ and $q=1$.
\newline
\textbf{Left diagram (ENED):} $\protect\beta=2$ (dashed line), $\protect\beta%
=4$ (doted line), $\protect\beta=6$ (continuous line) and Maxwell case (bold
line). \newline
\textbf{Right diagram (LNED):} $\protect\beta=1$ (dashed line), $\protect%
\beta=1.5$ (doted line), $\protect\beta=3$ (continuous line) and Maxwell
case (bold line).}
\label{Fig2}
\end{figure}
%%%%%%%%%%%%%%%%%%%%%%%%%%%%%%%%%%%%%%%%%%%%%%%%%%%%%%%%%%%%%%%%%%%%

%%%%%%%%%%%%%%%%%%%%%%%%%%%%%%%%%%%%%%%%%%%%%%%%%%%%%%%%%%%%%%%%%%%%
\begin{figure}[tbp]
$%
\begin{array}{ccc}
\epsfxsize=10cm \epsffile{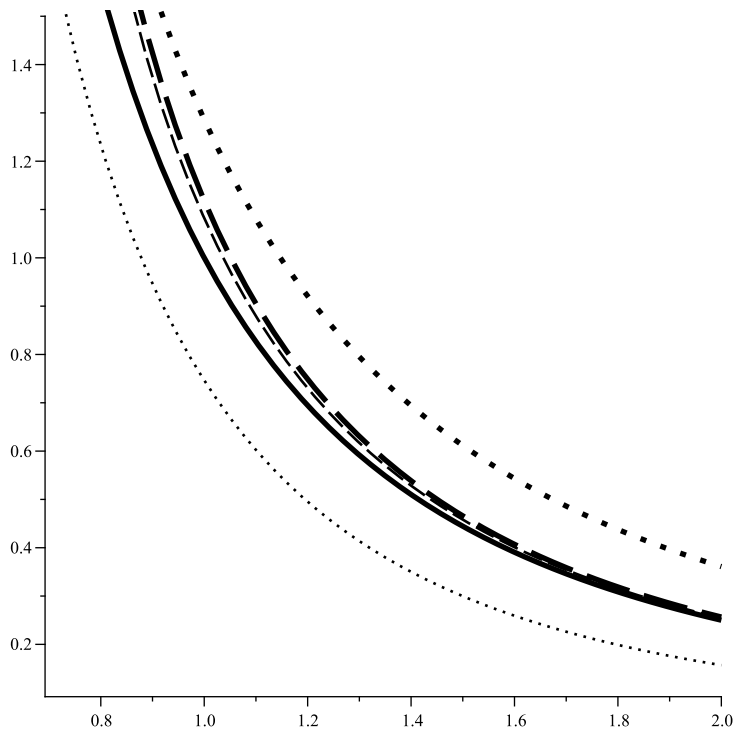} &  &
\end{array}
$%
\caption{\textbf{\emph{Comparison of various theories:}} $T^{_{\widehat{%
\protect\rho }\widehat{\protect\rho }}}$ versus $\protect\rho$ for $l=1$, $%
q=1$, $s=0.9$ (doted line for PMI), $s=1$ (bold continuous line for
Maxwell), $s=1.1$ (bold doted line for PMI), $\protect\beta=4$ (dashed line
for ENED) and $\protect\beta=4$ (bold dashed line for LNED).}
\label{Fig3}
\end{figure}
%%%%%%%%%%%%%%%%%%%%%%%%%%%%%%%%%%%%%%%%%%%%%%%%%%%%%%%%%%%%%%%%%%%%

As one can see, in case of PMI theory, we have a special case $s=1$ which is
denoted as Maxwell theory of electrodynamics. We consider two set of values
for $s$ in order to have a better understanding of the behavior of energy
density. These two cases are $\frac{1}{2}<s<1$ and $s>1$. It is evident
through studying these two cases (Fig. \ref{Fig1}) that as one increases $s$
parameter, the concentration of energy density increases. In other words,
for a fixed value of $\rho $ the lowest value of the energy density of
spacetime belongs to the lowest value of $s$. In general, the energy density
is a decreasing function of $\rho $.

Next for the two BI-types nonlinear electromagnetic fields we have plotted
Fig. \ref{Fig2} . Due to structure of their nonlinear electrodynamics, these
theories do not have real valued energy tensor every where. There is a
region in which their energy tensor is imaginary. This region is a
decreasing function of nonlinearity parameter. As one can see in case of
these two theories, increasing the nonlinearity parameter leads to
decreasing the concentration of energy density. In other words, in order to
decrease the concentration of energy density, one should increase the
nonlinearity parameter, $\beta$.

Finally we have plotted Fig. \ref{Fig3} in order to make a comparison
between these theories of electromagnetic fields. It is evident that the
lowest and highest value of energy density belongs to PMI theory. Regardless
of PMI case, the LNED has larger value of energy density and next one is
ENED.

\subsection{Geometric Properties}

Now we want to study the properties of the spacetime described by Eq. (\ref%
{metric}) with obtained metric functions of different NED models. First of
all, to investigate the singularities and asymptotical behavior of the
solutions, we calculate the Kretschmann scalar
\begin{equation}
\left. R_{\mu \nu \lambda \kappa }R^{\mu \nu \lambda \kappa }\right\vert
_{PMI}=12\Lambda ^{2}-4(4s-3)\Lambda \mathcal{D}+(8s^{2}-8s+3)\mathcal{D}%
^{2},  \label{PMIKretschmann}
\end{equation}%
\begin{equation}
\mathcal{D=}\left( \frac{\sqrt{2}q}{l\rho ^{1/(2s-1)}}\right) ^{2s},
\end{equation}%
\begin{equation}
\left. R_{\mu \nu \lambda \kappa }R^{\mu \nu \lambda \kappa }\right\vert
_{\exp }=12\Lambda ^{2}+4\Lambda (2\mathcal{J}+\mathcal{K})-(2\mathcal{J}%
^{2}+\mathcal{K}^{2}),  \label{EXPKretchmann}
\end{equation}%
\begin{equation}
\mathcal{J}=\beta ^{2}-\frac{2q\beta (L_{W}-1)}{\rho l\sqrt{-L_{W}}},
\end{equation}%
\begin{equation}
\mathcal{K}=\beta ^{2}+\frac{2q\beta }{\rho l\sqrt{-L_{W}}},
\end{equation}%
\begin{equation}
\left. R_{\mu \nu \lambda \kappa }R^{\mu \nu \lambda \kappa }\right\vert
_{\log }=12\Lambda ^{2}-96\Lambda \beta ^{2}\left( \Pi +\frac{2}{3}(1-\Gamma
)\right) +\mathcal{T},  \label{LOGKretchmann}
\end{equation}%
\begin{equation}
\mathcal{T}=192\beta ^{2}\left\{ \left( \ln \left[ 2q^{2}l^{2}\rho ^{2}\beta
^{2}\left( 1-\Gamma \right) \right] \right) ^{2}-\frac{4}{3}\left( 1-\Gamma
\right) \left[ 1-\ln \left[ l^{2}\rho ^{2}\beta ^{2}\left( 1-\Gamma \right) %
\right] \right] +\frac{4}{3}\left( 1-\frac{1}{l\rho \beta }\right) \ln
\left( 2q^{2}\right) -\frac{8q^{3}}{3l^{2}\rho ^{2}\beta ^{2}}\right\} .
\end{equation}

Regarding Eqs. (\ref{PMIKretschmann}), (\ref{EXPKretchmann}) and (\ref%
{LOGKretchmann}), it is easy to show that Kretschmann scalar diverges at $%
\rho =0$\ and therefore one might think that there is a curvature
singularity located at $\rho =0$; but as we will see, the spacetime will
never achieve $\rho =0$. Also, in the mentioned equations, the Kretschmann
scalar reduces to $12\Lambda ^{2}$ for $r\longrightarrow \infty $, which
confirms asymptotical behavior of these spacetimes is AdS. There are two
possible cases for the metric function. In one case, the metric function has
no root which is interpreted as naked singularity. In the other case, metric
function has one or more than one root. If one consider $r_{0}$ as the
largest root of metric function, there will be a change in signature of
metric. In other words, for $\rho <r_{0}$ metric function is negative, hence
metric signature is $(-,+,+)$. For $\rho >r_{0}$ metric function is
positive, therefore metric signature is $(-,-,-)$. This change in metric
signature results into a conclusion: it is not possible to extend spacetime
to $\rho <r_{0}$.

In order to cover the whole spacetime, correctly, we introduce
another coordinate transformation. The new radial coordinate $r$
may be introduces as
\begin{equation}
r^{2}=\rho ^{2}-r_{0}^{2}\Longrightarrow d\rho ^{2}=\frac{r^{2}}{%
r^{2}+r_{0}^{2}}dr^{2},  \label{coordinate}
\end{equation}%
where $\rho \geq r_{0}$ leads to $r\geq 0$. Applying this coordinate
transformation, the metric (\ref{metric}) should be written as
\begin{equation}
ds^{2}=-\frac{r^{2}+r_{0}^{2}}{l^{2}}dt^{2}+\frac{r^{2}}{\left(
r^{2}+r_{0}^{2}\right) g(r)}dr^{2}+l^{2}g(r)d\varphi ^{2},
\label{change coordinate metric}
\end{equation}%
where the coordinates $r$\ assumes the values $0\leq r<\infty $, and
obtained $g(r)$'s (Eqs. (\ref{PMI metric}), (\ref{ENEF metric}) and (\ref%
{LNEF metric})), are now given as%
\begin{equation}
\left. g(r)\right\vert _{PMI}=m-\Lambda \left( r^{2}+r_{0}^{2}\right)
+\left\{
\begin{array}{cc}
\frac{q^{2}\ln \left( \frac{r^{2}+r_{0}^{2}}{l^{2}}\right) }{l^{2}}, & s=1
\\
\frac{2^{s-1}(2s-1)^{2}}{\left( s-1\right) }\left( \frac{q}{l}\right)
^{2s}\left( r^{2}+r_{0}^{2}\right) ^{(s-1)/(2s-1)}, & otherwise%
\end{array}%
\right. ,  \label{change PMI metric}
\end{equation}

\begin{equation}
\left. g(r)\right\vert _{\exp }=m-\Lambda \left( r^{2}+r_{0}^{2}\right) -
\frac{\beta ^{2}\left( r^{2}+r_{0}^{2}\right) }{2}+\int \frac{4\beta qr}{l
\sqrt{r^{2}+r_{0}^{2}}}\left( \sqrt{-L_{W}^{\prime }}+\frac{1}{\sqrt{
-L_{W}^{\prime }}}\right) dr,  \label{change
ENEF metric}
\end{equation}

\begin{equation}
L_{W}^{\prime }=LambertW\left( -\frac{4q^{2}}{l^{2}\beta ^{2}\left(
r^{2}+r_{0}^{2}\right) }\right) ,
\end{equation}

\begin{equation}
\left. g(r)\right\vert _{\log }=m-\Lambda \left( r^{2}+r_{0}^{2}\right) +%
\frac{8q^{2}}{l^{2}}\ln \left[ l\beta \left( 1+\Gamma \right) \sqrt{%
r^{2}+r_{0}^{2}}\right] -2\left( r^{2}+r_{0}^{2}\right) \beta ^{2}\left[
3\left( 1-\Gamma \right) +2\Pi \right] ,  \label{change
LNEF metric}
\end{equation}%
\begin{equation}
\Gamma ^{^{\prime }}=\sqrt{1-\left( \frac{2q}{l\beta \sqrt{r^{2}+r_{0}^{2}}}%
\right) ^{2}},
\end{equation}%
\begin{equation}
\Pi ^{^{\prime }}=\ln \left( \frac{l^{2}\beta ^{2}\left(
r^{2}+r_{0}^{2}\right) (1-\Gamma ^{^{\prime }})}{2q^{2}}\right) .
\end{equation}

The nonzero component of electromagnetic field in the new coordinate can be
given by
\begin{equation}
\left. F_{\varphi r}\right\vert _{PMI}=q\left( r^{2}+r_{0}^{2}\right)
^{-1/(4s-2)},  \label{new PMI field}
\end{equation}%
\begin{equation}
\left. F_{\varphi r}\right\vert _{\exp }=\frac{l\beta }{2}\sqrt{%
-L_{W}^{\prime }},  \label{new ENEF field}
\end{equation}%
\begin{equation}
\left. F_{\varphi r}\right\vert _{\log }=\frac{l^{2}\beta ^{2}}{2q}\sqrt{%
r^{2}+r_{0}^{2}}\left( 1-\Gamma ^{^{\prime }}\right) .
\end{equation}

One can show that all curvature invariants do not diverge in the range $%
0\leq r<\infty $ and also $g(r)$, in different NED models namely Eqs. (\ref%
{change PMI metric}), (\ref{change ENEF metric}) and (\ref{change LNEF
metric}), is positive definite for $0\leq r<\infty $. It is evident through
studying the obtained values that in order to solutions contain singularity
both $r$ and $r_{0}$ must be zero whereas this case is never reached due to
considering nonzero value for $r_{0}.$ Therefore, these spacetimes have no
curvature singularity and horizon. However, the spacetime (\ref{change
coordinate metric}) has a conic geometry because the limit of the ratio
\emph{"circumference/radius"} is not $2\pi $ and therefore the spacetime has
a conical singularity at $r=0$
\begin{equation}
\lim_{r\longrightarrow 0}\frac{1}{r}\sqrt{\frac{g_{\varphi \varphi }}{g_{rr}}%
}\neq 1.
\end{equation}

The conical singularity can be removed if one exchanges the coordinate $%
\varphi $ with the following period
\begin{equation}
Period_{\varphi }=2\pi \left( \lim_{r\longrightarrow 0}\frac{1}{r}\sqrt{%
\frac{g_{\varphi \varphi }}{g_{rr}}}\right) ^{-1}=2\pi \left( 1-4\mu \right)
,  \label{Period}
\end{equation}%
where $\mu $ is given by%
\begin{equation}
\mu =\frac{1}{4}+\frac{1}{\Omega },  \label{miu}
\end{equation}%
where $\Omega $ is different for various models of NED. We find that%
\begin{equation}
\left. \Omega \right\vert _{PMI}=4lr_{0}\left[ \Lambda -\frac{(2s-1)}{2}%
\left( \frac{\sqrt{2}q}{lr_{0}^{1/(2s-1)}}\right) ^{2s}\right] ,
\label{omega}
\end{equation}%
where for $s=1$ this equation reduces to the Maxwell theory. In order to
have better understanding of the behavior of deficit angle, we calculate the
divergence points of the deficit angle in PMI model in which these points
are located at

\begin{equation}
\left. r_{0}\right\vert _{\delta \varphi \longrightarrow \infty }=\pm \left[
\frac{\left( 2s-1\right) }{2\Lambda }\left( \frac{\sqrt{2}q}{l}\right) ^{2s}%
\right] ^{\frac{2s-1}{2s}}.
\end{equation}

Due to complexity of obtained relation for deficit angle, it is not possible
to calculate roots of deficit angle analytically. Therefore we will study
them in context of graphs for deficit angle later. It is worthwhile to
mention that in case of Maxwell theory the divergency and roots of deficit
angle are obtained as follow

\begin{equation}
\left. r_{0}\right\vert _{\delta \varphi \longrightarrow \infty }=\pm \frac{q%
}{l\sqrt{\Lambda }},  \label{divMax1}
\end{equation}%
\begin{equation}
\left. r_{0}\right\vert _{\delta \varphi =0}=\frac{\pm \sqrt{1+4\Lambda q^{2}%
}-1}{2l\Lambda }.  \label{divMax2}
\end{equation}

As one can see, in case of Maxwell theory, the divergency is only seen in dS
spacetime. In other words, in AdS spacetime which is of our interest in this
paper, no divergency exists. In case of PMI theory, one can draw same
conclusion; no divergency is observed in AdS spacetime.

Next, using the method that was introduced for calculation of deficit angle,
in case of ENED, we have the following result
\begin{equation}
\left. \Omega \right\vert _{\exp }=\frac{q\mathcal{NE}\left( 2+\frac{q%
\mathcal{E}\sqrt{-L_{\mathcal{W}}}}{l\beta r_{0}}\right) +8q\beta ^{2}l^{2}%
\sqrt{-L_{\mathcal{W}}}(1+L_{\mathcal{W}})+\frac{l\beta r_{0}\mathcal{N}}{%
\sqrt{-L_{\mathcal{W}}}}}{\frac{\beta l^{2}r_{0}^{2}(1+L_{\mathcal{W}})}{8}%
\left( 2+\frac{2q\mathcal{E}\sqrt{-L_{\mathcal{W}}}}{l\beta r_{0}}\right)
^{2}},
\end{equation}%
where%
\begin{equation}
L_{\mathcal{W}}=LambertW\left( -\frac{4q^{2}}{l^{2}\beta ^{2}r_{0}^{2}}%
\right) ,
\end{equation}%
\begin{equation}
\mathcal{N}=2ql\beta r_{0}L_{\mathcal{W}}^{2}+\sqrt{-L_{\mathcal{W}}}\left[
l^{2}r_{0}^{2}(1+L_{\mathcal{W}})(2\Lambda +\beta ^{2})+4q^{2}\exp \left( -%
\frac{1}{2}L_{\mathcal{W}}\right) \right] ,
\end{equation}%
\begin{equation}
\mathcal{E}={Ei}\left( 1,\frac{1}{2}L_{\mathcal{W}}\right) ,
\end{equation}%
and the exponential integral, ${Ei}\left( a,z \right) $, are defined for ${Re%
}(z)>0$ (${Re}(z)$ means the real part of $z$) by
\begin{equation}
{Ei}\left( a,z\right) =\int_{1}^{\infty }\exp \left( -zx\right) x^{-a}dx.
\end{equation}

It is notable to mention that, in order to have a real deficit angle, $%
\delta \phi $, we should consider $\beta >\beta _{\min }$, where
\begin{equation}
\beta _{\min }=\frac{2q}{lr_{0}}\exp \left( \frac{1}{2}\right) .
\label{betaminexp}
\end{equation}

By applying same process for the case of LNED, one can find
\begin{equation}
\left. \Omega \right\vert _{\log }=16\beta ^{2}lr_{0}\left\{ 1+\frac{\Lambda
}{4\beta ^{2}}-\mathcal{H}+\ln \left[ -\frac{2}{\beta (1-\mathcal{H})}\left(
\frac{ql}{r_{0}}\right) ^{2}\right] \right\} ,
\end{equation}%
\begin{equation}
\mathcal{H=}\sqrt{1+\left( \frac{2ql}{\beta r_{0}}\right) ^{2}}.
\end{equation}

For logarithmic and exponential forms, due to complexity of obtained deficit
angle relations, it is not possible to find singular points of deficit
angle, analytically. But by employing numerical method, it was seen that the
singularity is located at $r_{0}=0$, which we should note that, the system
never reaches this limit.

\begin{figure}[tbp]
$%
\begin{array}{ccc}
\epsfxsize=8cm \epsffile{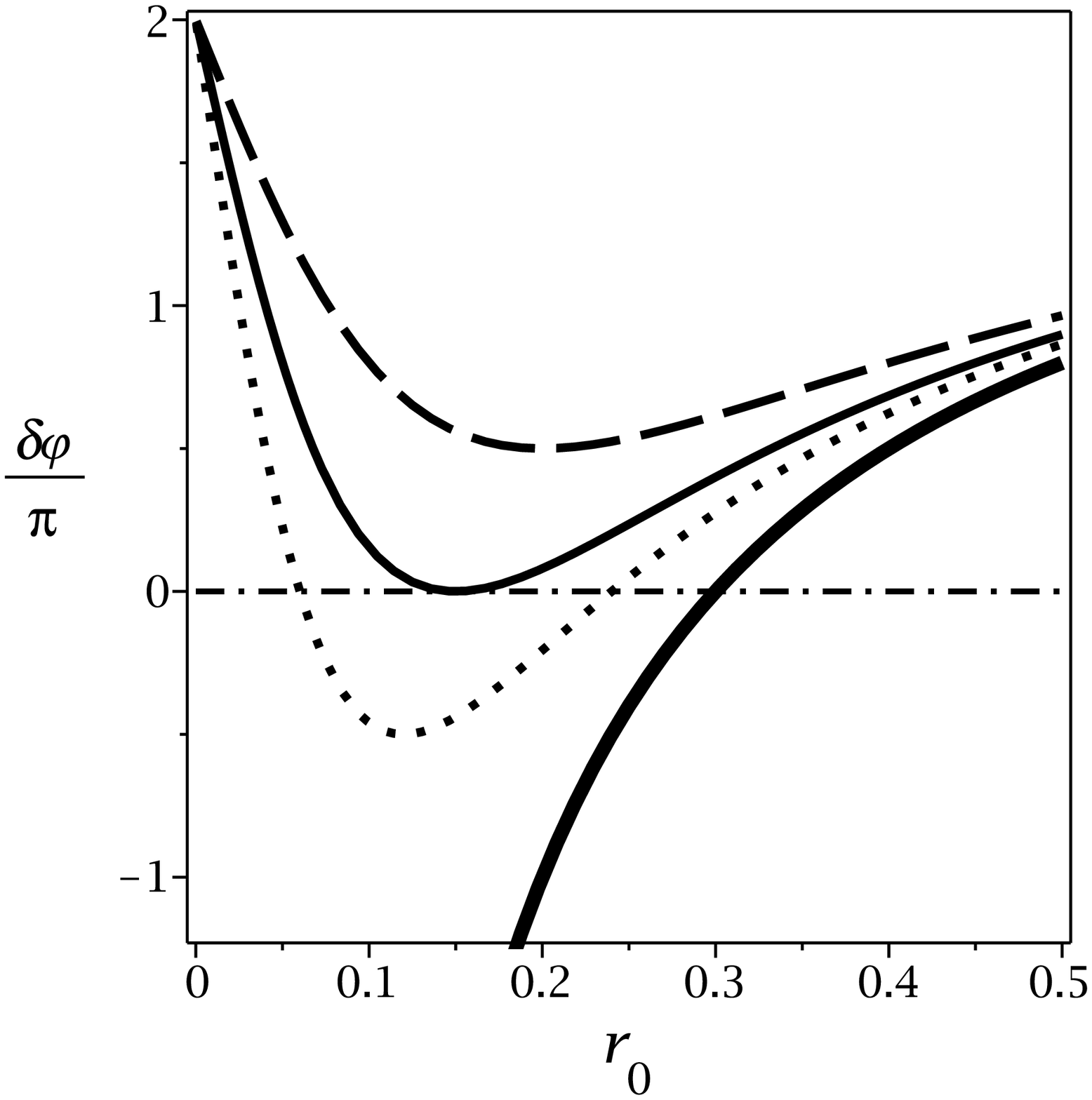} & \epsfxsize=8cm %
\epsffile{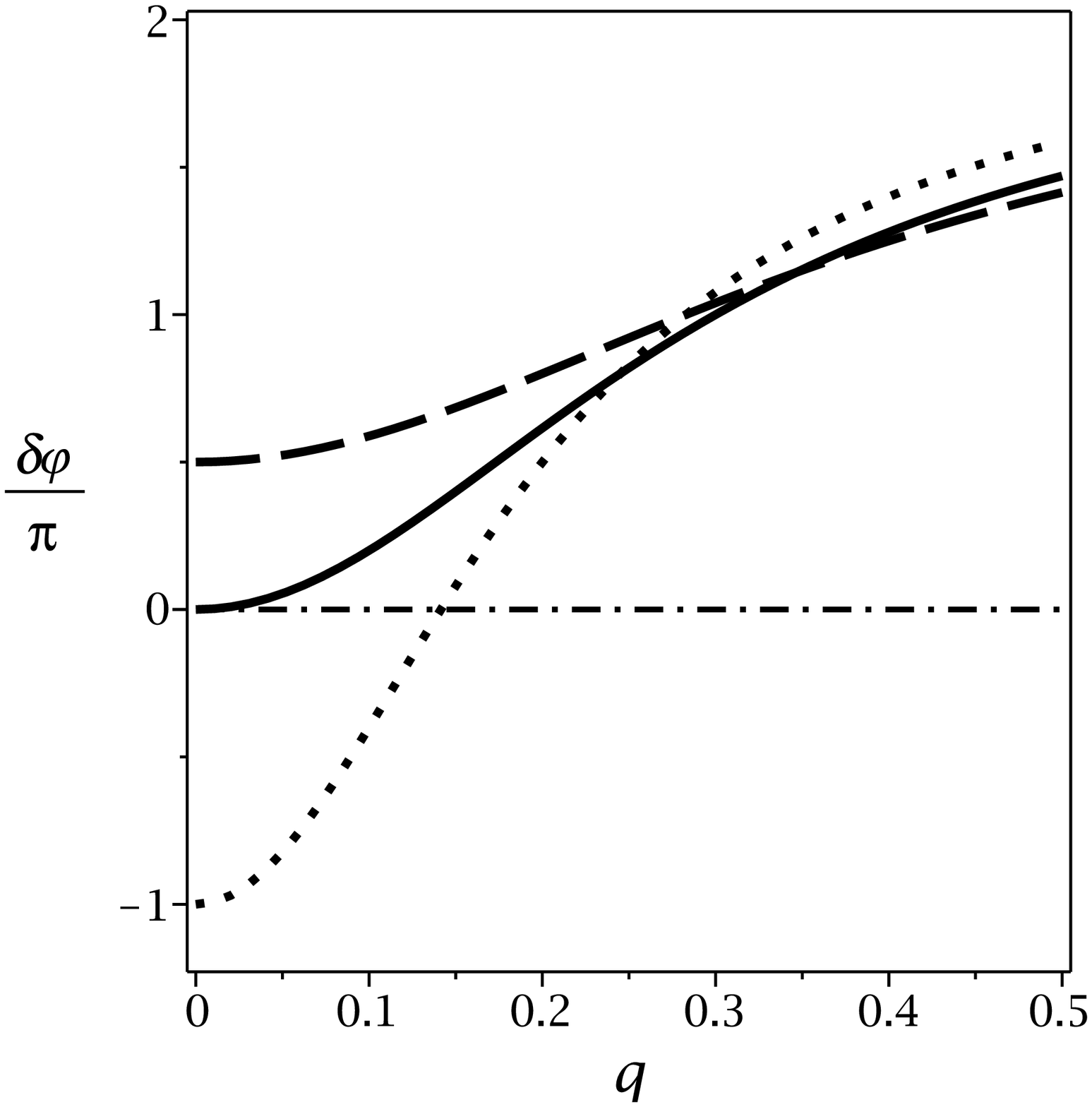} &
\end{array}
$%
\caption{\textbf{\emph{Maxwell solutions:}} $\protect\delta
\protect\phi$/$\protect\pi$ versus $r_{0}$ (left) and
$\protect\delta \protect\phi$/$\protect\pi$ versus $q$ (right) for
$l=0.3$.
\newline \textbf{Left diagram:} $q=0$ (bold line), $q=0.12$ (doted
line), $q=0.15$ (continuous line) and $q=0.2$ (dashed line).
\newline \textbf{Right diagram:} $r_{0}=0.2$ (doted line),
$r_{0}=0.3$ (continuous line) and $r_{0}=0.4$ (dashed line).}
\label{Fig33}
\end{figure}

\begin{figure}[tbp]
$%
\begin{array}{ccc}
\epsfxsize=8cm \epsffile{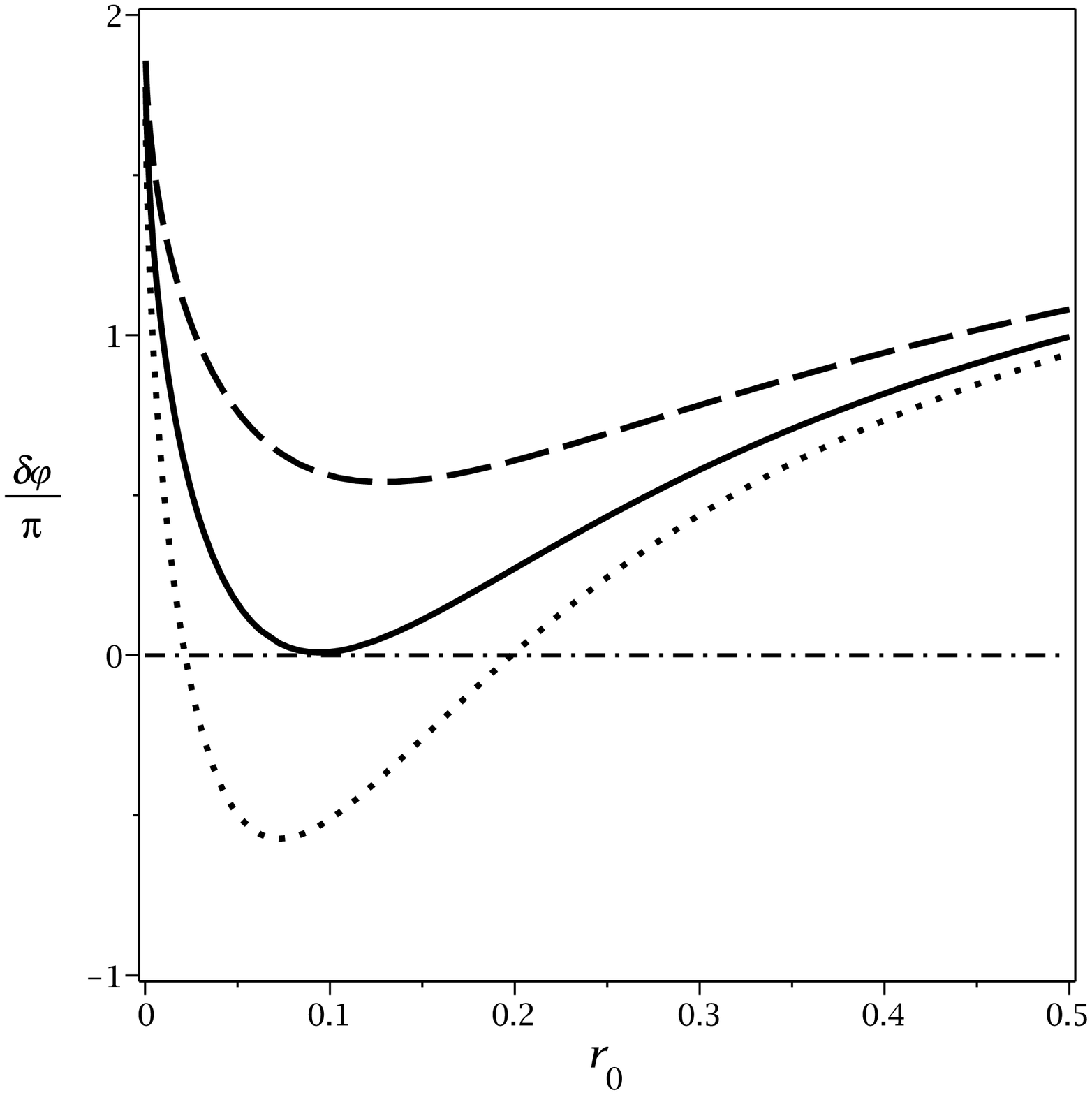} & \epsfxsize=8cm %
\epsffile{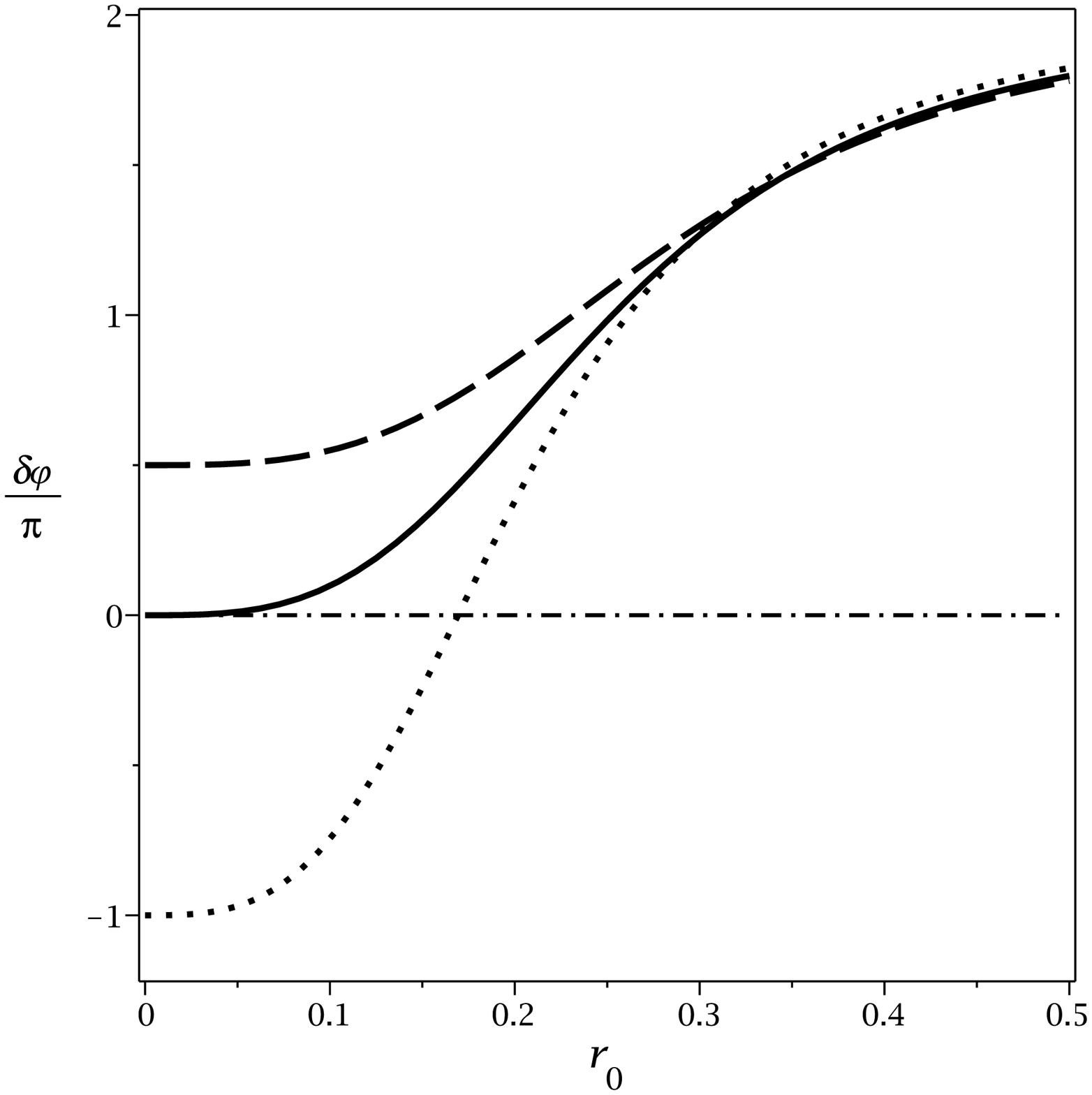} &
\end{array}
$%
\caption{\textbf{\emph{PMI solutions:}} $\protect\delta
\protect\phi$/$\protect\pi$ versus $r_{0}$ (left) and
$\protect\delta \protect\phi$/$\protect\pi$ versus $q$ (right) for
$l=0.3$ and $s=1.6$. \newline \textbf{Left diagram:} $q=0.17$
(doted line), $q=0.191$ (continuous line) and $q=0.22$ (dashed
line).
\newline \textbf{Right diagram:} $r_{0}=0.2$ (doted line),
$r_{0}=0.3$ (continuous line) and $r_{0}=0.4$ (dashed line).}
\label{Fig4}
\end{figure}

\begin{figure}[tbp]
$%
\begin{array}{ccc}
\epsfxsize=5cm \epsffile{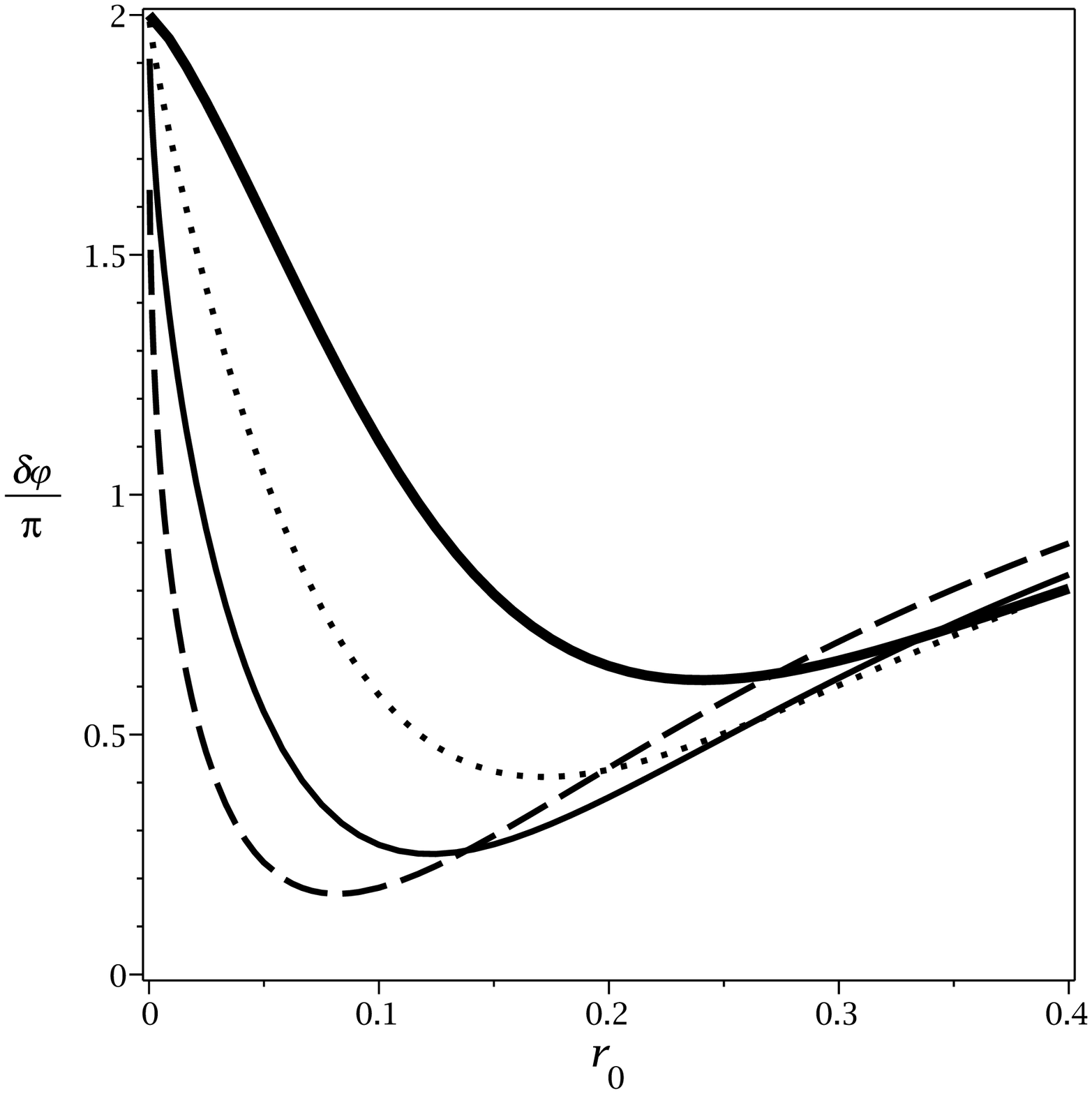} & \epsfxsize=5cm %
\epsffile{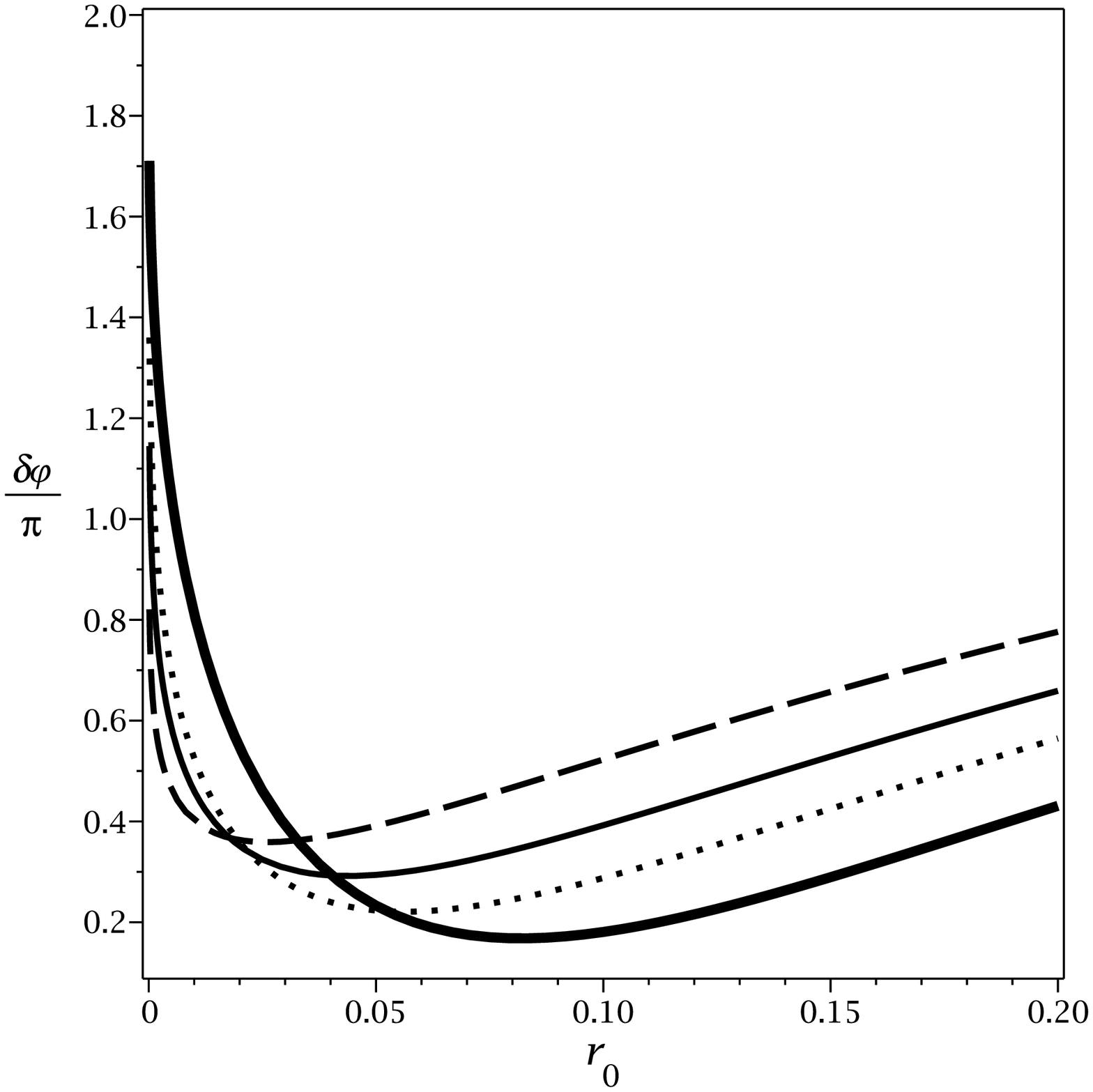} & \epsfxsize=5cm \epsffile{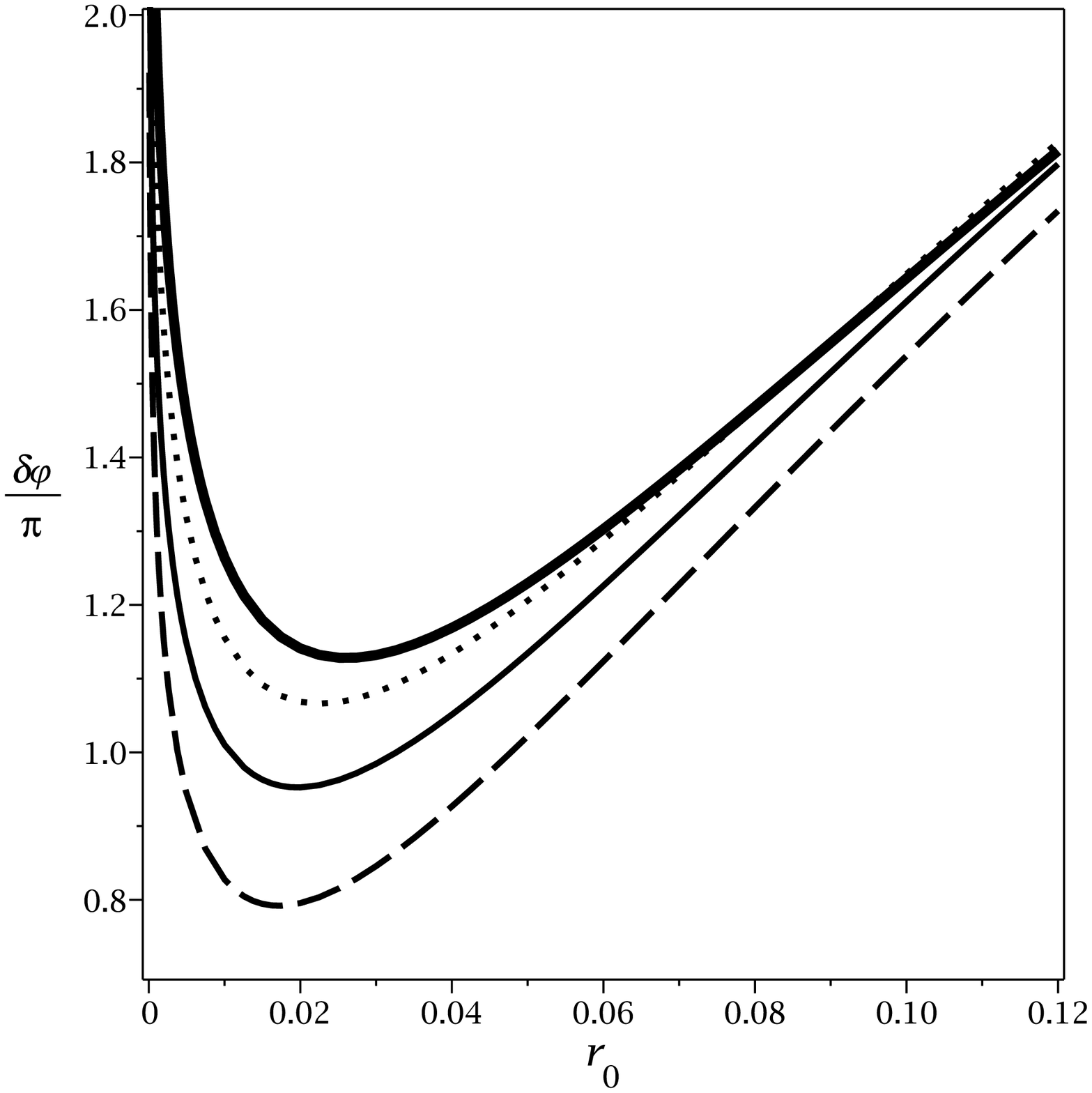}%
\end{array}
$%
\caption{\textbf{\emph{PMI solutions:}} $\protect\delta
\protect\phi$/$\protect\pi$ versus $r_{0} $ for $l=0.3$ and
$q=0.2$. \newline \textbf{Left diagram:} $s=0.9$ (bold line),
$s=1.1$ (doted line), $s=1.4$ (continuous line) and $s=2$ (dashed
line). \newline \textbf{Middle diagram:} $s=2$ (bold line), $s=3$
(doted line), $s=4$ (continuous line) and $s=7$ (dashed line).
\newline \textbf{Right diagram:} $s=7$ (bold line), $s=8$ (doted
line), $s=9$ (continuous line) and $s=10$ (dashed line).}
\label{Fig5}
\end{figure}

\begin{figure}[tbp]
$%
\begin{array}{ccc}
\epsfxsize=5.5cm \epsffile{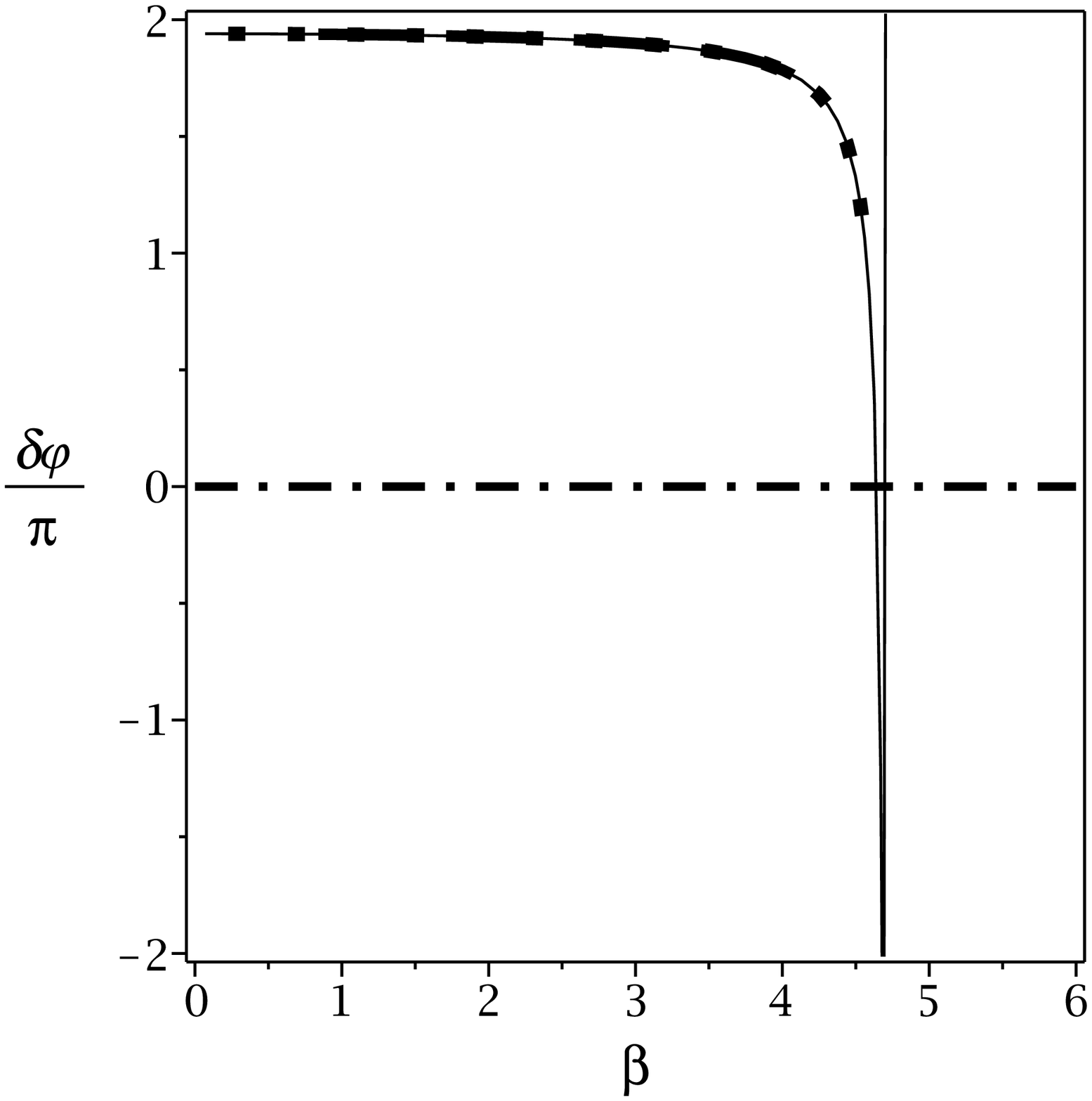} & \epsfxsize=5.5cm %
\epsffile{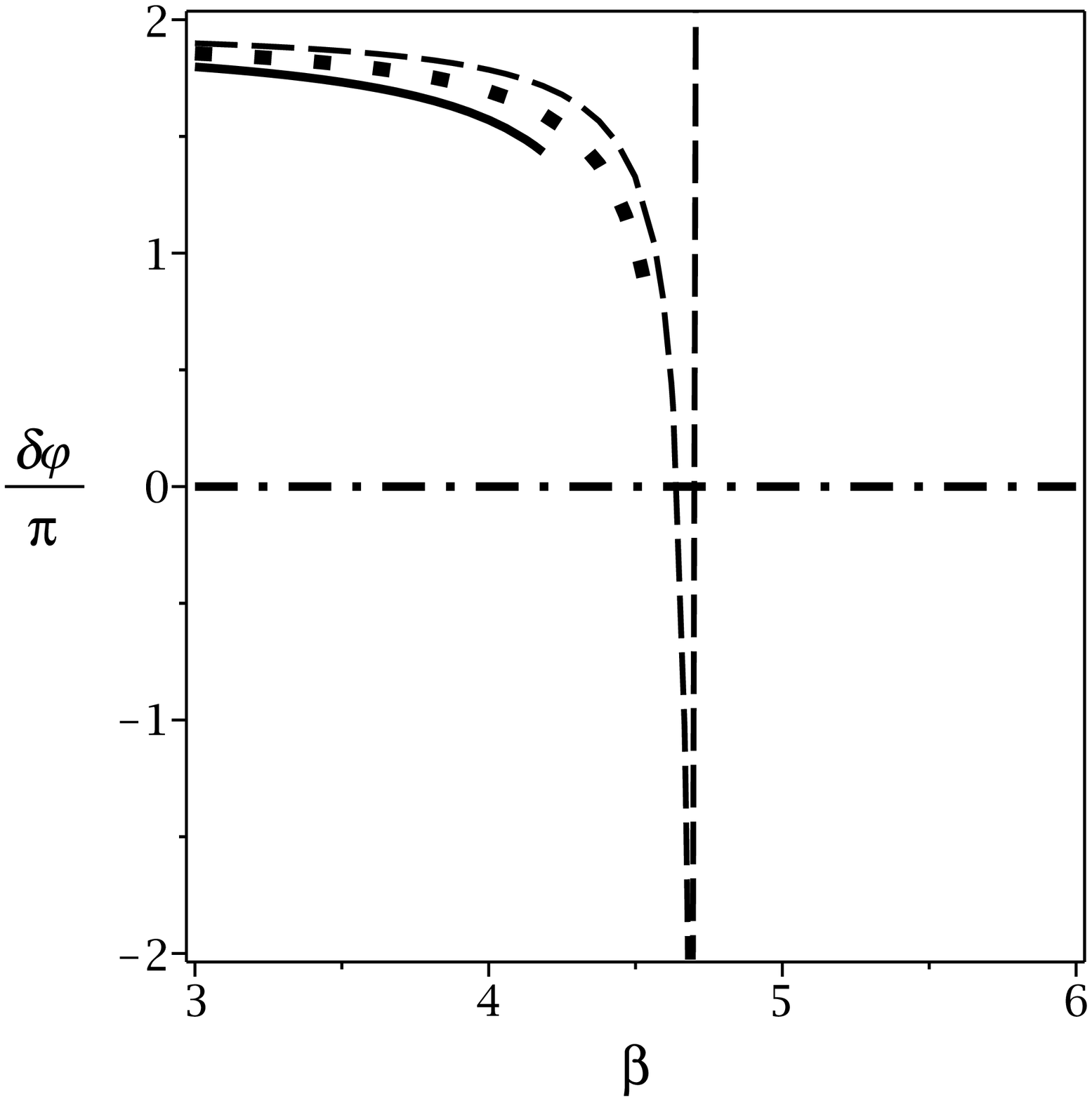} & \epsfxsize=5.5cm %
\epsffile{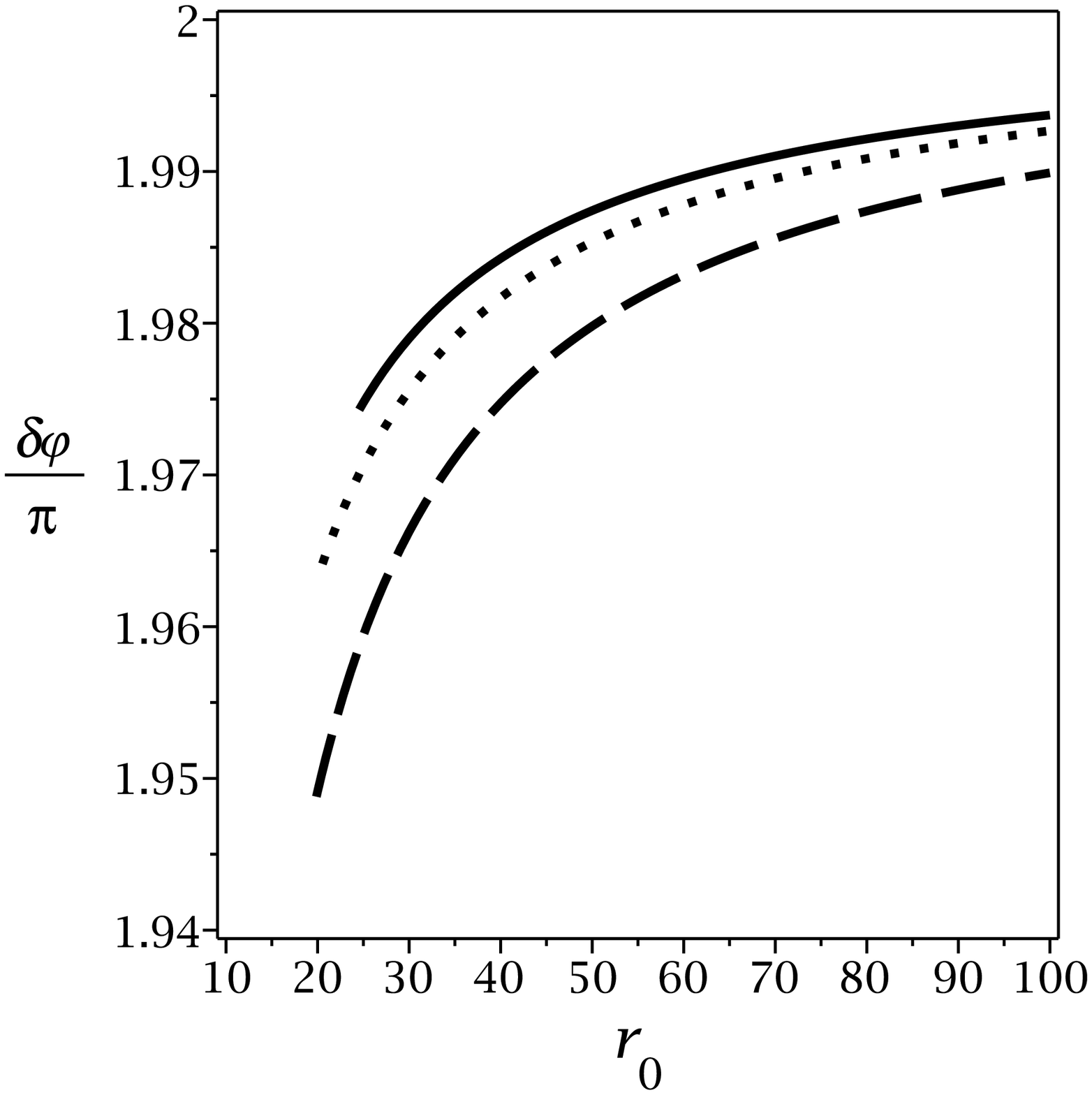}%
\end{array}
$%
\caption{\textbf{\emph{ENED solutions:}} $\protect\delta
\protect\phi$/$\protect\pi$ versus $\protect\beta$ (left and
middle) and $\protect\delta \protect\phi$/$\protect\pi$ versus
$r_{0}$ (right) for $l=0.3$ \newline \textbf{Left diagram:}
$r_{0}=10$, $q=0.06$ (continuous line), $q=0.1$ (doted line) and
$q=0.2$ (dashed line).
\newline \textbf{Middle diagram:} $q=0.06$, $r_{0}=5$ (continuous
line), $r_{0}=7$ (doted line) and $r_{0}=10$ (dashed line).
\newline
\textbf{Right diagram:} $q=1$, $\protect\beta =1$ (continuous line), $%
\protect\beta =2$ (doted line) and $\protect\beta =3$ (dashed
line).} \label{Fig6}
\end{figure}

\begin{figure}[tbp]
$%
\begin{array}{ccc}
\epsfxsize=5.5cm \epsffile{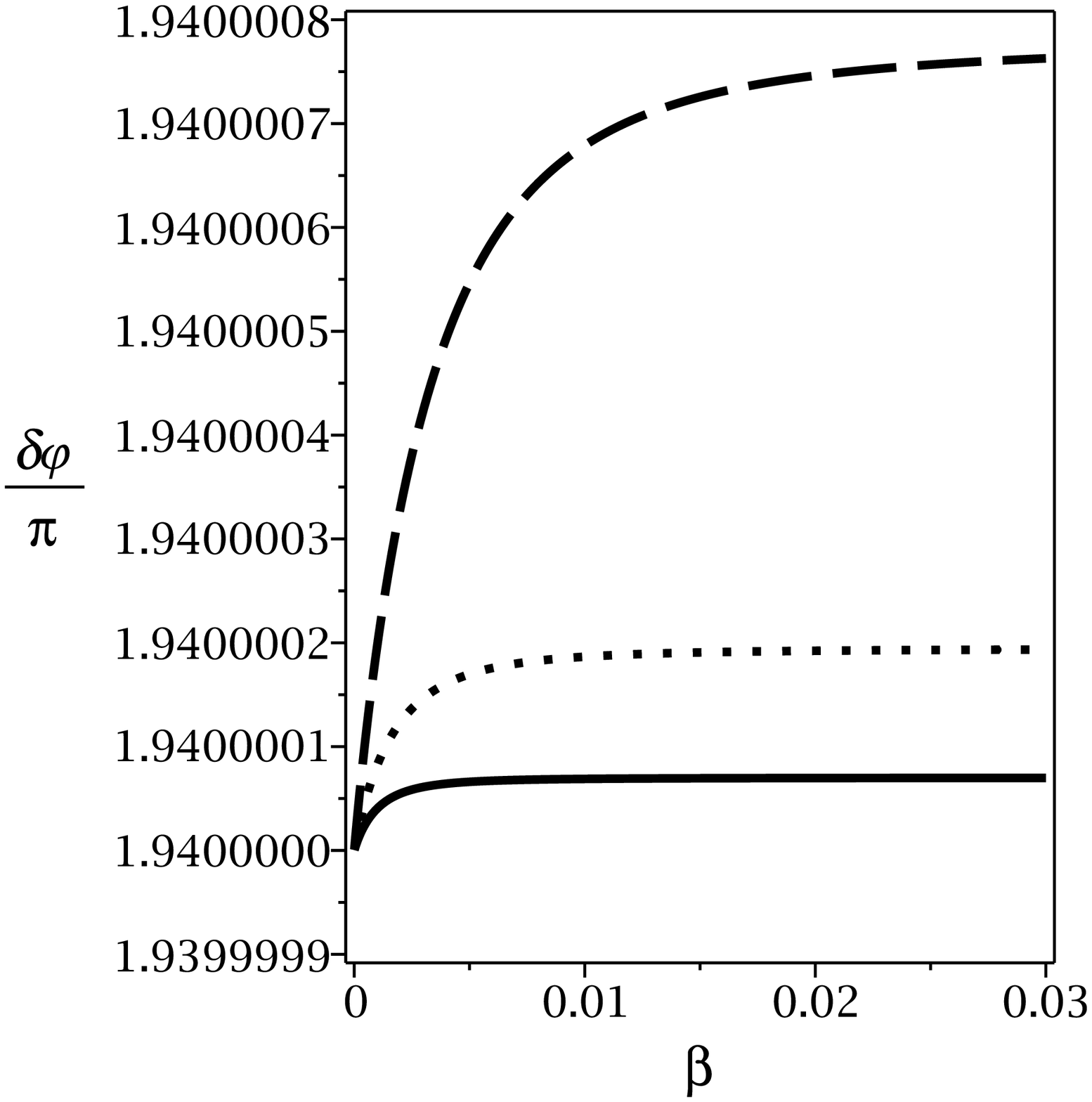} & \epsfxsize=5.5cm %
\epsffile{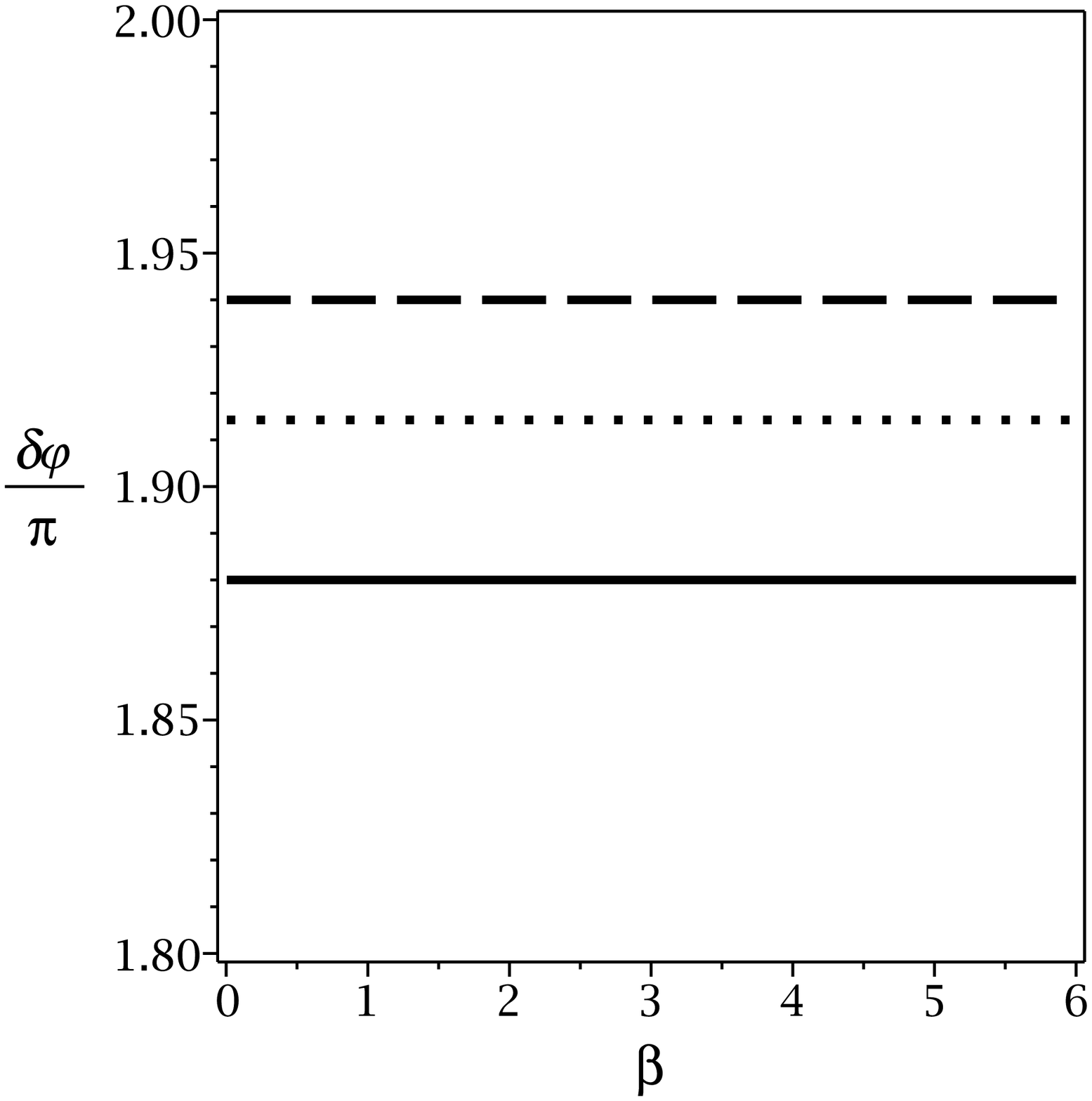} & \epsfxsize=5.5cm %
\epsffile{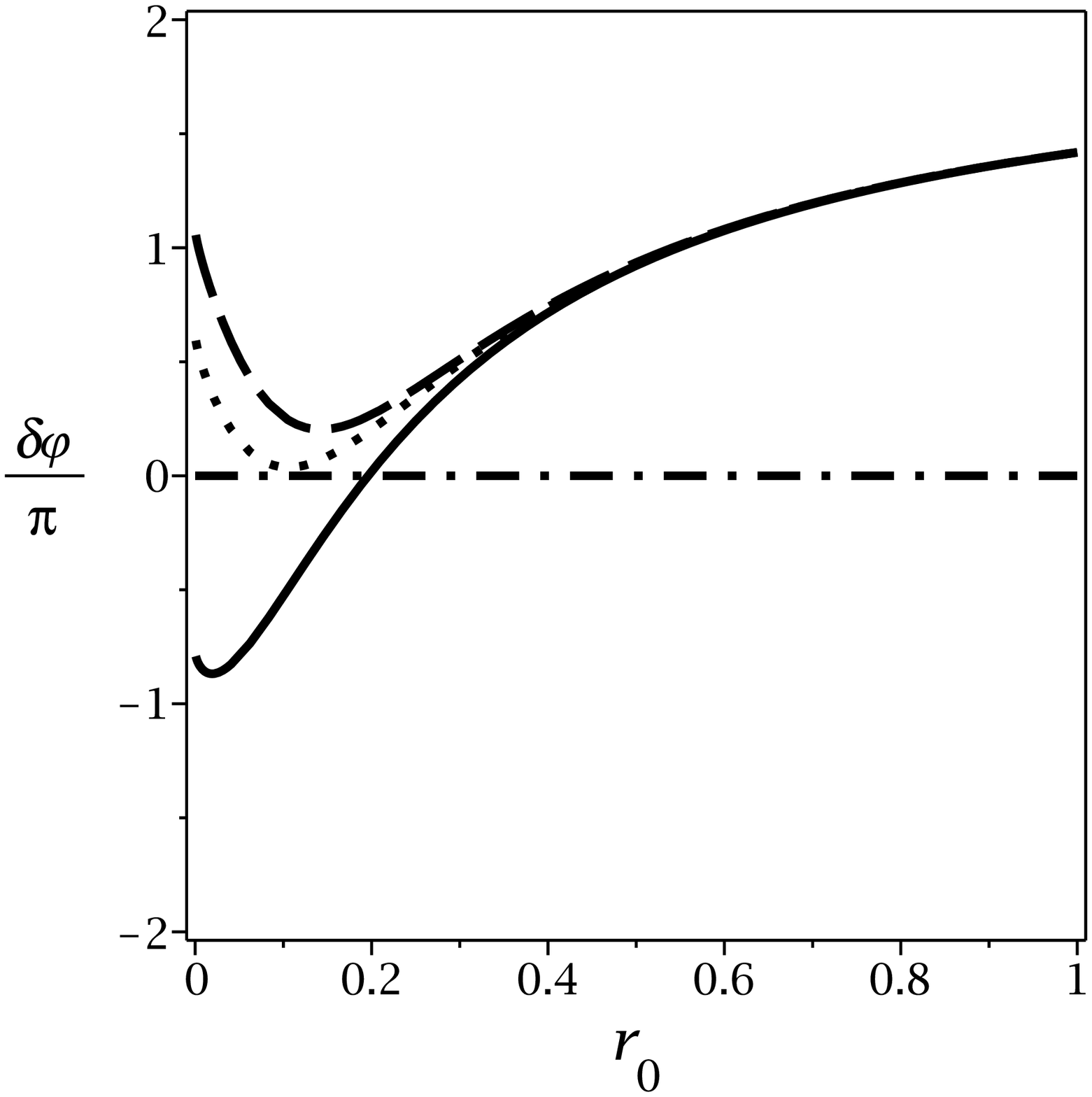}%
\end{array}
$%
\caption{\textbf{\emph{LNED solutions:}} $\protect\delta
\protect\phi$/$\protect\pi$ versus $\protect\beta$ (left and
middle) and $\protect\delta \protect\phi$/$\protect\pi$ versus
$r_{0}$ (right) for $l=0.3$ \newline \textbf{Left diagram:}
$r_{0}=10$, $q=0.06$ (continuous line), $q=0.1$ (doted line) and
$q=0.2$ (dashed line).
\newline \textbf{Middle diagram:} $q=0.06$, $r_{0}=5$ (continuous
line), $r_{0}=7$ (doted line) and $r_{0}=10$ (dashed line).
\newline
\textbf{Right diagram:} $q=1$, $\protect\beta =1$ (continuous line), $%
\protect\beta =2$ (doted line) and $\protect\beta =3$ (dashed
line).} \label{Fig7}
\end{figure}

Now, we are in position to study the effects of variation of different
parameters on the deficit angle in these nonlinear theories. To do so, we
have plotted Figs. (\ref{Fig33})-(\ref{Fig7}).

In case of PMI theory, obtained results are as follow. Due to structure of
equation of deficit angle in the presence of this nonlinear electromagnetic
field, small values of $s$ $\left( 0<s<1/2\right) $ are considered non
physical. Therefore, we will only consider large values of $s$ $\left(
s>1/2\right) $. As for the case $s=1$, the PMI theory will reduce to Maxwell
theory, before we study nonlinear theories, we first investigate the
properties of the Maxwell solutions.

To investigate the effects of charge on the deficit angle, we plot Fig. \ref%
{Fig33} (left). In this case, there is a minimum for deficit angle located
at $r_{0_{\min }}$. This $r_{0_{\min }}$ and corresponding deficit angle to
it are increasing function of charge. In the other words, there is a
critical value of charge, $q_{c}$, which for $q<q_{c}$, deficit angle has
two roots and a region of negativity whereas for $q>q_{c}$, deficit angle
has no root and always positive. It is worthwhile to mention that in absence
of charge, deficit angle is only increasing function of $r_{0}$.

In case of varying $r_{0}$\ (Fig. \ref{Fig33} right), deficit angle is an
increasing function of charge. In this case, there also exists a critical
value, $r_{0_{c}}$in which for $r_{0}<r_{0_{c}}$, there will be a region of
negativity and a root for calculated values of deficit angle. For case of $%
r_{0}>r_{0_{c}}$, deficit angle is positive and without any root. It is
worthwhile to mention that for small values of charge the highest values of
deficit angle belongs to highest value of $r_{0}.$ As charge increases large
enough, this behavior will change; the highest values of the deficit angle
belongs to the lowest value of $r_{0}$.

As in case of PMI theory for variation of charge Fig. \ref{Fig4} (left) is
plotted. For this case, there is a minimum $r_{0_{\min }}$ and the deficit
angle corresponding to $r_{0}=r_{0_{\min }}$ are increasing functions of
charge. There is a critical value for charge, $q_{c}$ in which deficit angle
corresponds to it, is zero. For $q<q_{c}$, there will be two roots for
deficit angle, otherwise deficit angle does not have root. Next, in order to
study the effects of variation of $r_{0}$, we have Fig. \ref{Fig4} (right).
The behavior of deficit angle and the effects of varying $r_{0}$ is similar
to the case of varying $r_{0}$ in the Maxwell theory.

As for the effects of $s$, we plot Fig. \ref{Fig5}. Interestingly, different
behaviors are seen for different values of $s$. The general behavior of the
graphs are similar to variation of charge. But with one unique property.
There is a different behavior for plotted graphs of deficit angle. First the
$r_{0_{\min }}$ and corresponding deficit angle to it are decreasing
functions of $s$ (Fig. \ref{Fig5} left). This behavior will change as one
increases $s$ which results into $r_{0_{\min }}$ being decreasing function
of $s$ whereas deficit angle corresponding to $r_{0}=r_{0_{\min }}$ is an
increasing function of $s$ (Fig. \ref{Fig5} middle). This behavior will
change again if one increases $s$ which leads to $r_{0_{\min }}$ and deficit
angle of it being decreasing function of $s$ (Fig. \ref{Fig5} right).

As for exponential form of nonlinear electromagnetic field, it is
seen that due to its structure, there is a divergency for deficit
angle for the cases of the deficit angle versus nonlinearity
parameter (Figs. \ref{Fig6} left and middle). It is evident that
the existence of the divergency is a function of the variations of
charge and $r_{0}$. In other words, for sufficiently small (large)
values of charge ($r_{0}$), there will be a divergency for deficit
angle. Whereas, by increasing (decreasing) charge ($r_{0}$)
instead of divergency, there will be a region in which deficit
angle is not real. This region is an increasing (decreasing)
function of the charge ($r_{0}$) (Fig. \ref{Fig6} left and
middle).

Next, for the effects of nonlinearity parameter on deficit angle
(Fig. \ref{Fig6} right), the region where deficit angle is not
real is seen in this case too. It is evident that this region is a
decreasing function of nonlinearity parameter. Overall, the
deficit angle in this case is an increasing function of the
$r_{0}$ whereas for the case of the nonlinearity, it is a
decreasing function of the $\beta$. This shows that effects of the
these two parameters on deficit angle are opposite of each other.

\begin{figure}[tbp]
$%
\begin{array}{ccc}
\epsfxsize=5.5cm \epsffile{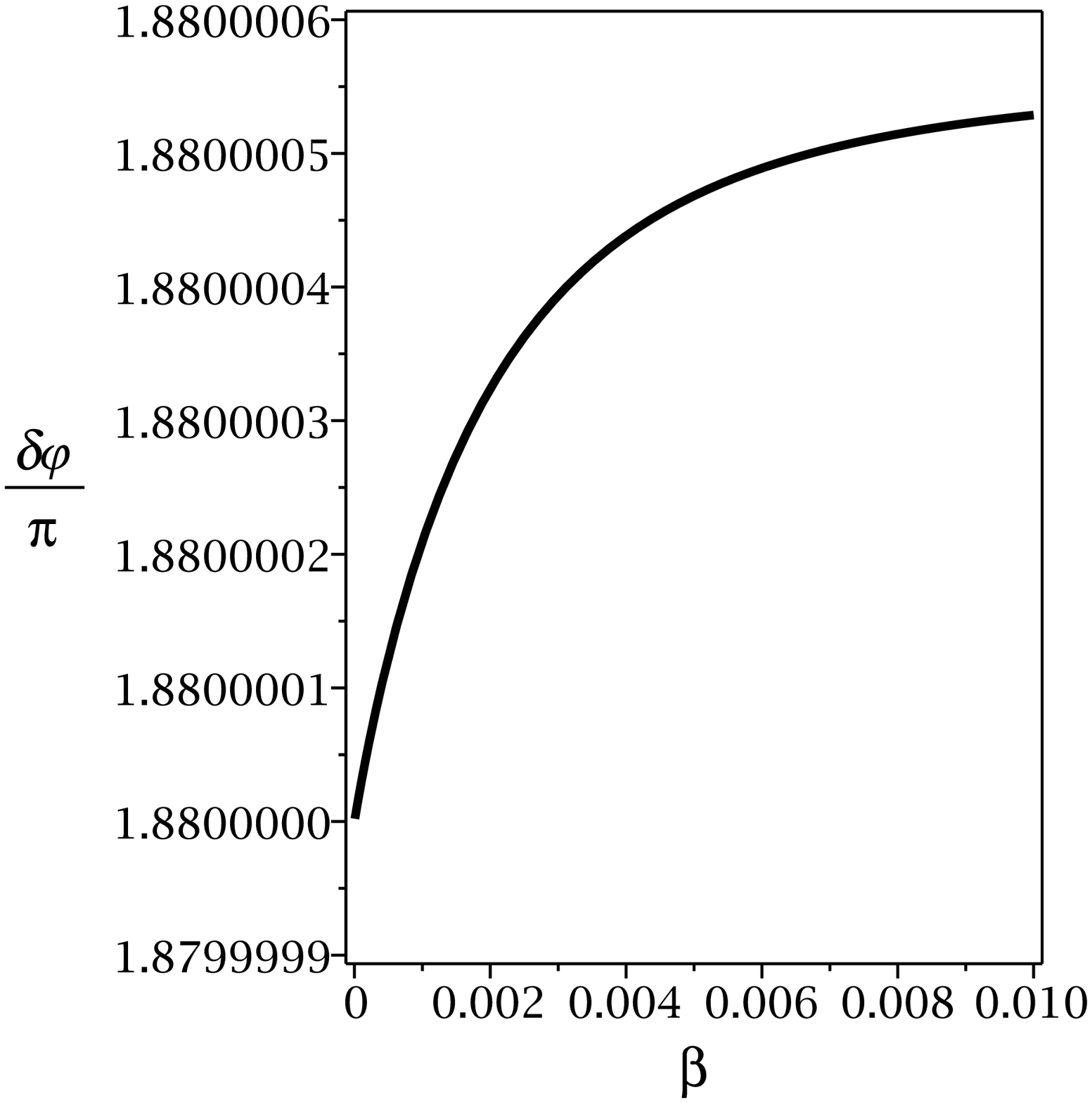} & \epsfxsize=5.5cm %
\epsffile{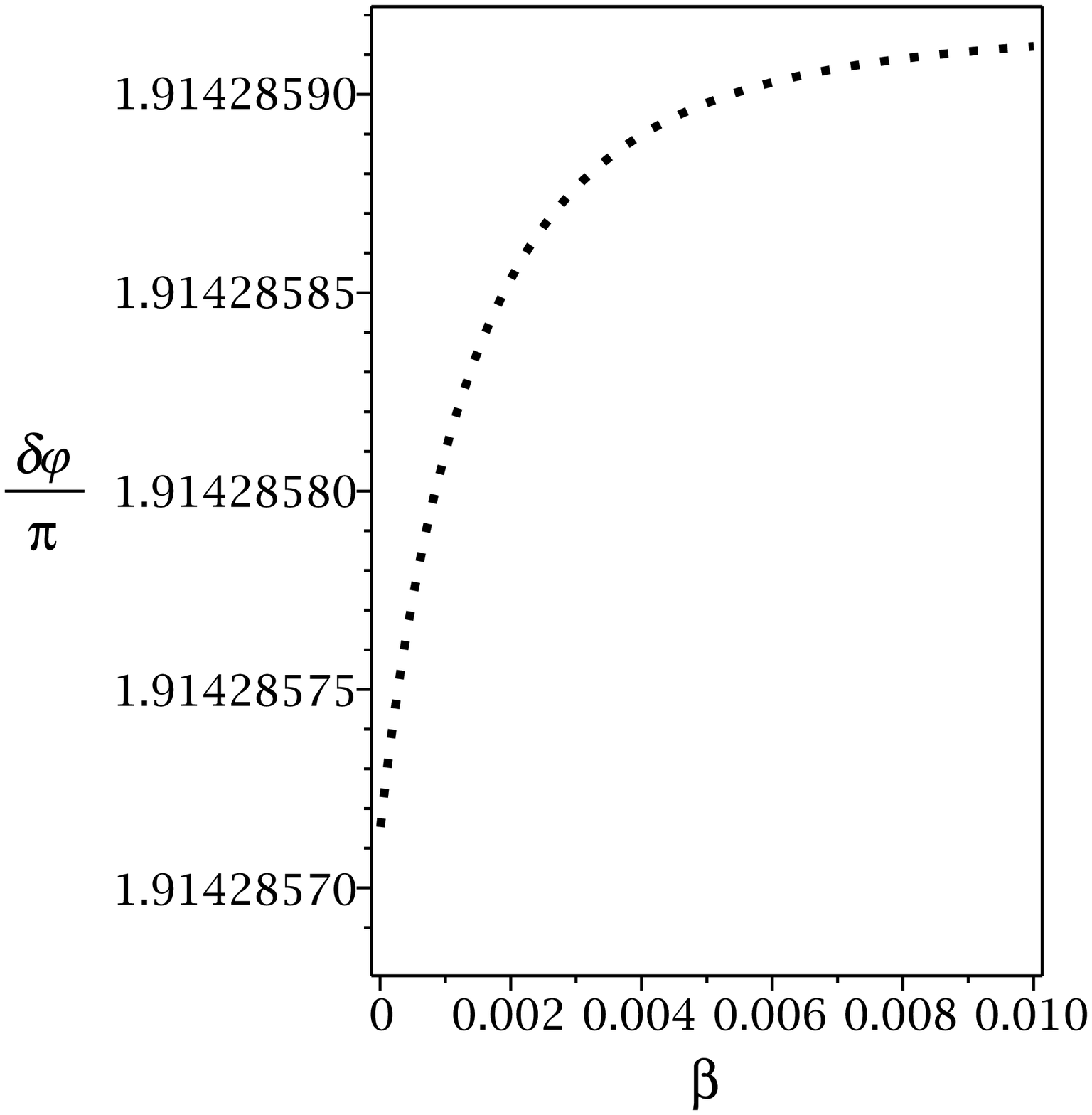} & \epsfxsize=5.5cm %
\epsffile{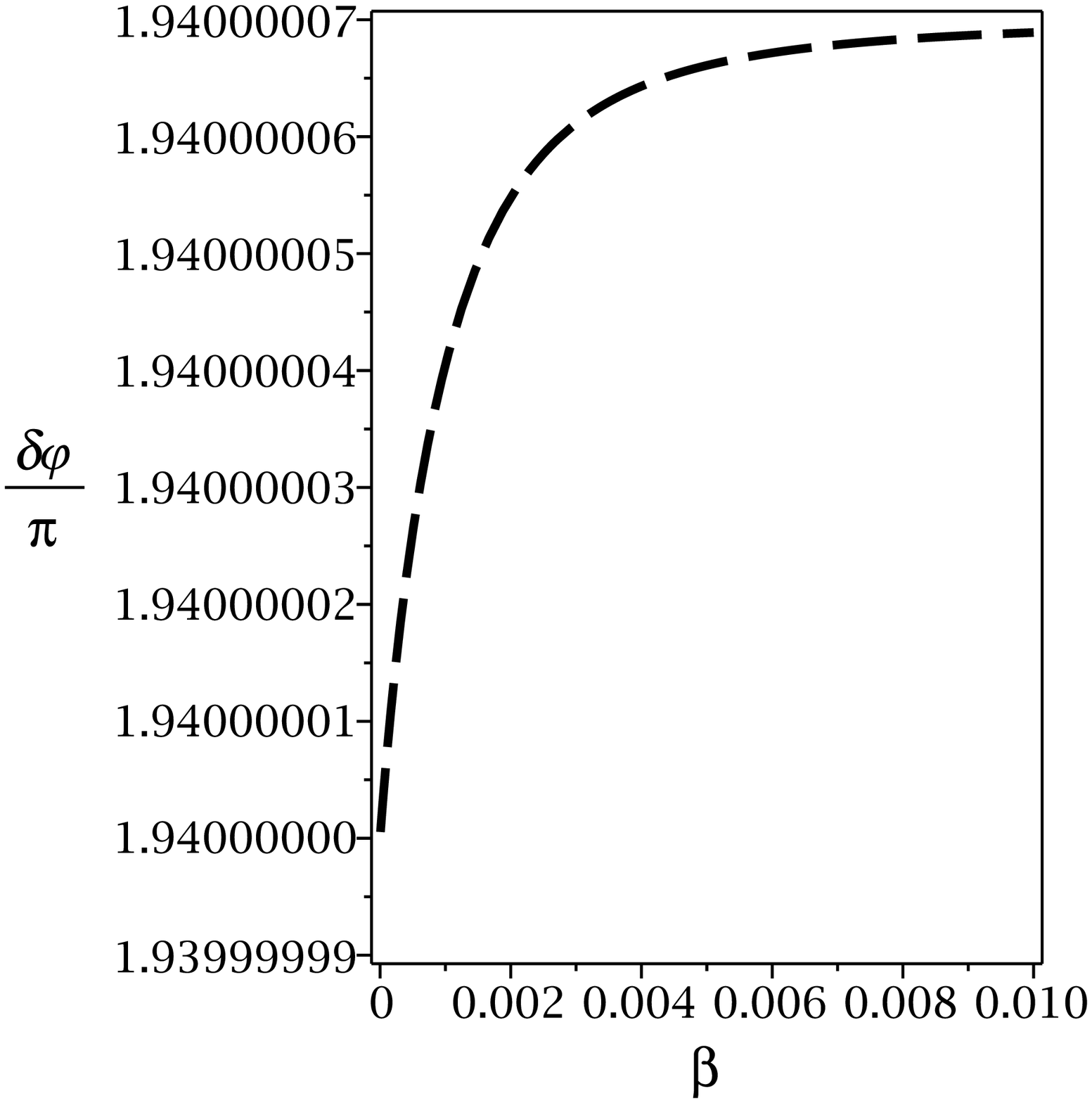}%
\end{array}
$%
\caption{\textbf{\emph{LNED solutions:}} Different scales of
 $\protect\delta \protect\phi$/$\protect\pi$ versus $\protect\beta$ for
$q=0.06$, and $l=0.3$, $r_{0}=5$ (continuous line), $r_{0}=7$
(doted line) and $r_{0}=10$ (dashed line).} \label{Fig77}
\end{figure}

For the case of logarithmic nonlinear electromagnetic field, one
can find following results. As for the effects of charge, plotted
graph (Fig. \ref{Fig7} left) shows that deficit angle is an
increasing function of nonlinearity parameter and for case of
fixing nonlinearity and other parameters, the highest value of
deficit angle belongs to the highest value of charge. On the other
hand, as for the effects of $r_{0}$ (Figs. \ref{Fig7} middle and
\ref{Fig77}), the calculated values of deficit angle are positive
and for $\beta =0$ deficit angle is also positive and non zero. As
$r_{0}$ decreases, the related value of deficit angle for case of
$\beta =0$, decreases. In general, in this case too, the deficit
angle is an increasing function of the nonlinearity parameter. In
order to showing this behavior, we have plotted Fig. \ref{Fig77}.

For the effects of nonlinearity parameter, Fig. \ref{Fig7} (right)
is plotted. As one can see, interestingly, the behavior of this
graph is quite different comparing to previous case. Deficit angle
is positive and non zero for case of $r_{0}=0$. The value of
deficit angle for this case is an increasing function of
nonlinearity parameter. Remarkably, two behaviors for deficit
angle are seen for this case and no singularity takes place. There
is an extremum $r_{0_{ext}}$in which for case of $r_{0}\leq
r_{0_{ext}}$, deficit angle is a decreasing function of $r_{0}$
and for case of $r_{0}\geq
r_{0_{ext}}$, it is an increasing function of $r_{0}$. For large values of $%
r_{0}$ the effect of nonlinearity will decrease and obtained
values of deficit angle for different cases of nonlinearity
parameter will be so close. From Fig. \ref{Fig7} one can show that
the highest and lowest values, in logarithmic form of nonlinear
electrodynamics, of deficit angle is located at
\begin{equation}
\left. \delta \varphi \right\vert _{\min }=\lim_{\beta \rightarrow 0}\delta
\varphi ,
\end{equation}%
\begin{equation}
\left. \delta \varphi \right\vert _{\max }=\lim_{\beta \rightarrow \infty
}\delta \varphi .
\end{equation}

%%%%%%%%%%%%%%%%%%%%%%%%%%%%%%%%%%%%%%%%%%%%%%%%%%%%%%%%%%%%%%%%%%%%
%%%%%%%%%%%%%%%%%%%%%%%%%%%%%%%%%%%%%%%%%%%%%%%%%%%%%%%%%%%%%%%%%%%%
%%%%%%%%%%%%%%%%%%%%%%%%%%%%%%%%%%%%%%%%%%%%%%%%%%%%%%%%%%%%%%%%%%%

%%%%%%%%%%%%%%%%%%%%%%%%%%%%%%%%%%%%%%%%%%%%%%%%%%%%%%%%%%%%%%%%%%%%%%
%%%%%%%%%%%%%%%%%%%%%%%%%%%%%%%%%%%%%%%%%%%%%%%%%%%%%%%%%%%%%%%%%%%%%%

%%%%%%%%%%%%%%%%%%%%%%%%%%%%%%%%%%%%%%%%%%%%%%%%%%%%%%%%%%%%%%%%%%%%%%

\subsection{$\emph{Spining~Solutions}$}

Now, we would like to endow our spacetime solution (\ref{metric}) with a
global rotation. In order to\ add an angular momentum to the spacetime, we
perform the following local rotation boost in the $t-\varphi $ plane
\begin{equation}
\begin{array}{cc}
t\longmapsto \Xi t-a\varphi , & \varphi \longmapsto \Xi \varphi -\frac{a}{%
l^{2}}t,%
\end{array}
\label{rotating boost}
\end{equation}%
where $a$ is the rotation parameter and $\Xi =\sqrt{1+a^{2}/l^{2}}$.
Inserting Eq. (\ref{rotating boost}) into Eq. (\ref{change coordinate metric}%
)\ we obtain
\begin{equation}
ds^{2}=-\frac{r^{2}+r_{0}^{2}}{l^{2}}\left( \Xi dt-ad\varphi \right) ^{2}+%
\frac{r^{2}}{\left( r^{2}+r_{0}^{2}\right) g(r)}dr^{2}+l^{2}g(r)\left( \frac{%
a}{l^{2}}dt-\Xi d\varphi \right) ^{2},  \label{rotating metric}
\end{equation}%
where $g(r)$\ is the same as that given in Eqs. (\ref{change PMI metric}), (%
\ref{change ENEF metric}) and (\ref{change LNEF metric}) for different NEDs.
The nonzero components of the electromagnetic field are given as
\begin{equation}
F_{rt}=-\frac{a}{\Xi l^{2}}F_{r\varphi },  \label{rotating PMI
field}
\end{equation}%
where Eq. (\ref{rotating PMI field}) is valid for all mentioned models.

The local transformation (\ref{rotating boost}) generates a new metric,
because it is not a proper coordinate transformation on the entire manifold.
Therefore, the metric (\ref{change coordinate metric}) and (\ref{rotating
metric}) can be locally mapped into each other but not globally, and so they
are distinct. It is notable to mention that $g(r)$ is always positive for $%
r>0$ and this spacetime has a conical singularity at $r=0$.

Now, we want to obtain the electric charge of the solutions. To determine
the electric field, we should consider the projection of the electromagnetic
field tensor on special hypersurface. The electric charge can be found by
calculating the flux of the electric field at infinity, yielding%
\begin{equation}
Q=\frac{q}{2\pi }\sqrt{\Xi ^{2}-1}\times \left\{
\begin{array}{cc}
2^{s-1}sq^{2s-2}, & PMI \\
1, & LNED\text{, }ENED%
\end{array}%
\right.
\end{equation}

Notice that the electric charge is proportional to the rotation parameter,
and is zero for the static spacetime. Finally, we calculate the conserved
quantities of these solutions. The mass and the angular momentum of the
magnetic solution can be calculated through the use of counterterm method.
Using Eqs. (\ref{M}) and (\ref{J}), and the suitable counterterm Lagrangian $%
\mathcal{L}_{ct}=1/l$, one finds
\begin{equation}
M=\frac{m}{8}\left( 2\Xi ^{2}-1\right) ,  \label{MPMI}
\end{equation}%
\begin{equation}
J=\frac{\Xi ma}{4},  \label{JPMI}
\end{equation}
which Eq. (\ref{JPMI}) confirms that $a$ is rotation parameter.

\section{Nonlinearity as a Correction}

It is arguable that instead of considering a theory which has the property
of being highly nonlinear, one can add correction terms to the Maxwell
theory in which we define as additional correction (AC). Regarding the weak
field limit of nonlinear models, one can add quadratic Maxwell invariants to
the Lagrangian of Maxwell theory. This consideration can be justified
through following reasons. First of all, in series expanding BI types of
nonlinear theories, the first leading term, which is coupled with
nonlinearity parameter, is quadratic Maxwell invariant. Second, in low
energy effective of string theory, this term could be obtained which gives
strong motivation for considering this kind of modification. Third, in order
to find experimental result for nonlinear theories one should take into
account only small degrees of nonlinearity. Therefore, one can build another
nonlinear theory of electromagnetic field \cite{Hendi9}.

The BI-type Lagrangians (\ref{ENED}) and (\ref{LNED}) for the large values
of $\beta $\ ($\beta $ is nonlinear parameter) tend to the following
nonlinear Lagrangian
\begin{equation}
\mathcal{L}_{AC}(\mathcal{F})=-\mathcal{F}+\alpha \mathcal{F}^{2}+O\left(
\alpha ^{2}\right) ,  \label{Lagrangian}
\end{equation}%
where $\alpha $ is nonlinearity parameter and is proportional to the inverse
value of nonlinearity parameter in BI-type theories, namely $\beta $. In Eq.
(\ref{Lagrangian}), the nonlinearity parameter is small and so the effects
of this parameter should be considered as a perturbation and in the limit $%
\alpha \longrightarrow 0$, $\mathcal{L}_{AC}(\mathcal{F})$ reduces to the
Lagrangian of the standard Maxwell field, $\mathcal{L}_{Maxwell}(\mathcal{F}%
)=-\mathcal{F}$.

In this case, we want to obtain the solutions of Einstein gravity in
presence of the nonlinear electrodynamics, which presented by the Lagrangian
(\ref{Lagrangian}), for static and rotating metrics. As described in the
previous sections, considering Eqs. (\ref{Maxwell equation}), (\ref{metric})
and (\ref{Lagrangian}), one can show that
\begin{equation}
\left( 1-\frac{12\alpha F_{\varphi \rho }^{2}}{l^{2}}\right) F_{\varphi \rho
}^{\prime }+\left( 1+\frac{4\alpha F_{\varphi \rho }^{2}}{l^{2}}\right)
\frac{F_{\varphi \rho }}{\rho }+O(\rho^2)=0,  \label{Diff}
\end{equation}%
where Eq. (\ref{Diff})\ has the following solution%
\begin{equation}
F_{\varphi \rho }=\frac{q}{\rho }-\frac{4q^{3}\alpha }{\rho ^{3}l^{2}}%
+O\left( \alpha ^{2}\right) .  \label{Lagrangian field}
\end{equation}

\subsubsection{Static AC Magnetic Solution}

To obtain the function $g\left( \rho \right) $, one can insert Eqs. (\ref%
{metric}) and (\ref{Lagrangian field}) in the gravitational field equation (%
\ref{Field equation}) to obtain \ the metric function $g(\rho )$ as%
\begin{equation}
g(\rho )=m-\Lambda \rho ^{2}+\frac{2q^{2}}{l^{2}}\ln \left( \frac{\rho }{l}%
\right) +\frac{2q^{4}\alpha }{l^{4}\rho ^{2}}+O\left( \alpha ^{2}\right) ,
\label{metricfun}
\end{equation}%
where $m$ is the integration constant which is related to the mass of
solutions. One can show that the metric (\ref{metric}) with the metric
function (\ref{metricfun})\ has a singularity at $\rho =0$ by calculating
the Kretschmann scalar as
\begin{equation}
R_{\mu \nu \lambda \kappa }R^{\mu \nu \lambda \kappa }=12\Lambda ^{2}-\frac{%
8\Lambda q^{2}}{l^{2}\rho ^{2}}-\frac{4q^{4}(4\Lambda \alpha -3)}{l^{4}\rho
^{4}}-\frac{80q^{6}\alpha }{l^{6}\rho ^{6}}+O\left( \alpha ^{2}\right) .
\label{kre}
\end{equation}

From Eq. (\ref{kre}) it is obvious that Kretschmann scalar divergence at $%
\rho =0$ and reduces to $12\Lambda ^{2}$\ for $\rho \longrightarrow \infty $%
. On the other hand, as mentioned before, because of changing in signature,
it is not possible to extend spacetime to $\rho <r_{0}$. Also, one can apply
the coordinate transformation (\ref{coordinate}) to the metric (\ref{metric}%
)\ and find the metric function as
\begin{equation}
g(\rho )=m-\Lambda \left( r^{2}+r_{0}^{2}\right) +\frac{2q^{2}}{l^{2}}\ln
\left( \frac{\left( r^{2}+r_{0}^{2}\right) ^{1/2}}{l}\right) +\frac{%
2q^{4}\alpha }{l^{4}\left( r^{2}+r_{0}^{2}\right) }+O\left( \alpha
^{2}\right) ,  \label{change Lagrangian metric}
\end{equation}%
\ and the electromagnetic field in the new coordinate is
\begin{equation}
F_{\varphi r}=\frac{q}{\left( r^{2}+r_{0}^{2}\right) ^{1/2}}-\frac{%
4q^{3}\alpha }{\left( r^{2}+r_{0}^{2}\right) ^{3/2}l^{2}}+O\left( \alpha
^{2}\right) .
\end{equation}

Since all curvature invariants do not diverge in the range $0\leq r<\infty $%
, one finds that there is no essential singularity. But, like previous
cases, this spacetime has a conical singularity at $r=0$ with the deficit
angle $\delta \varphi =8\pi \mu $ where $\mu $ is given by Eq. (\ref{miu})
and $\Omega $ has the following form
\begin{equation}
\left. \Omega \right\vert _{AC}=4lr_{0}\left[ \Lambda -\left( \frac{q}{lr_{0}%
}\right) ^{2}+2\alpha \left( \frac{q}{lr_{0}}\right) ^{4}\right] +O\left(
\alpha ^{2}\right) .
\end{equation}

\begin{figure}[tbp]
$%
\begin{array}{ccc}
\epsfxsize=5.5cm \epsffile{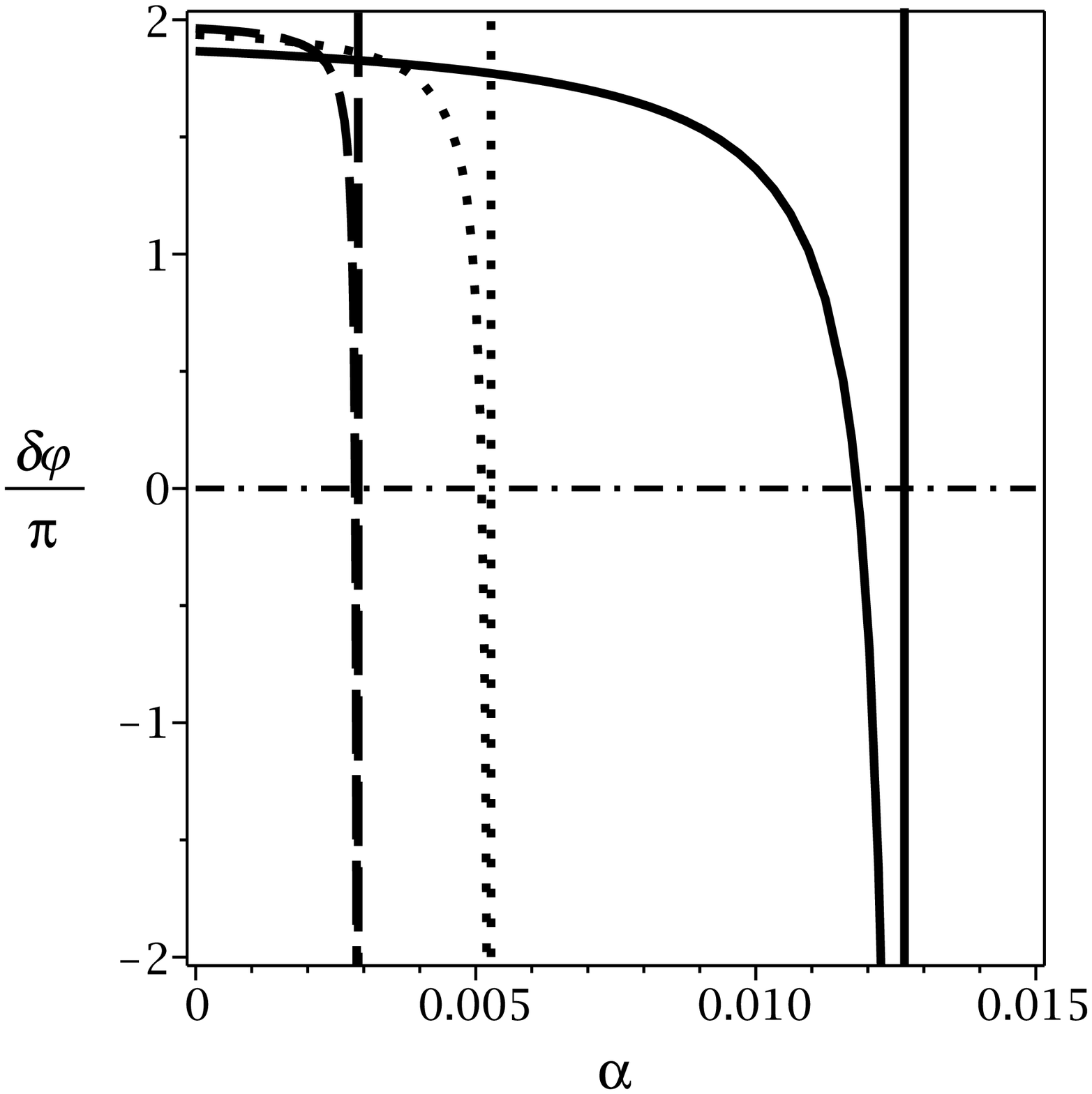} & \epsfxsize=5.5cm %
\epsffile{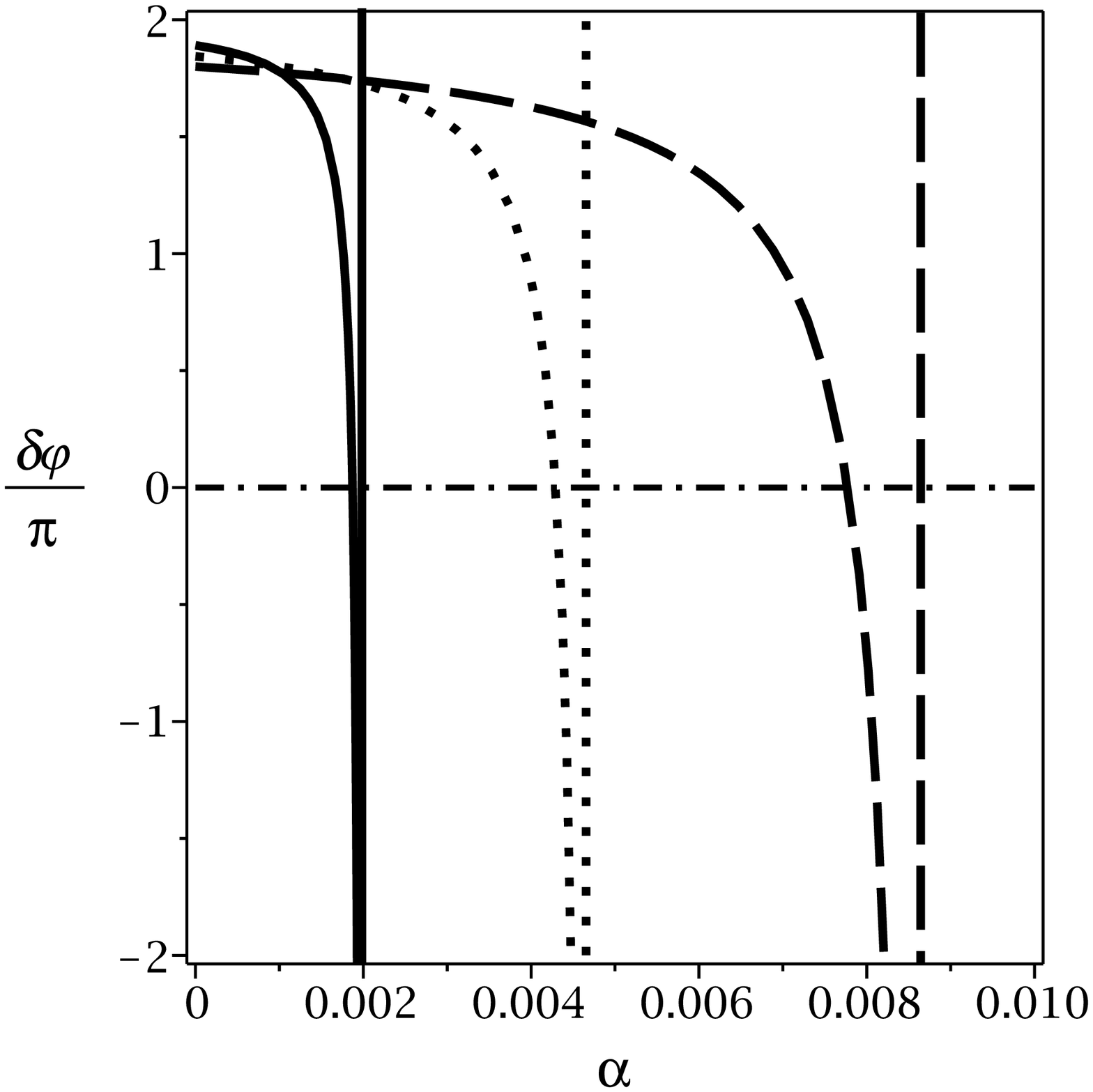} & \epsfxsize=5.5cm %
\epsffile{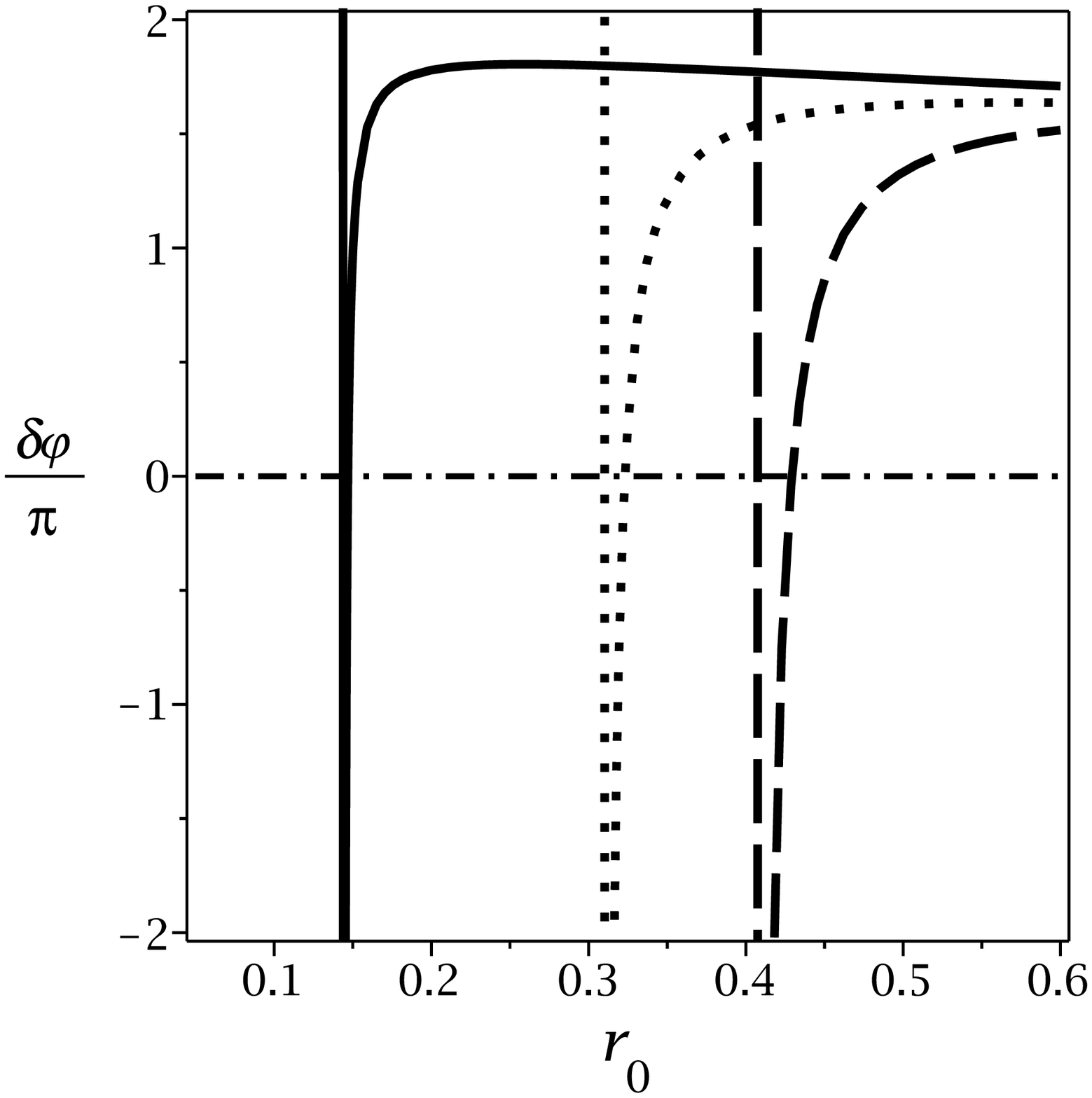}%
\end{array}
$%
\caption{\textbf{\emph{AC solutions:}} $\protect\delta
\protect\phi$/$\protect\pi$ versus $\protect\alpha$ (left and
middle) and $\protect\delta \protect\phi$/$\protect\pi$ versus
$r_{0} $ (right) for $l=0.3$. \newline \textbf{Left diagram:}
$r_{0}=1$, $q=2$ (continuous line), $q=3$ (doted line) and $q=4$
(dashed line). \newline \textbf{Middle diagram:} $q=1$,
$r_{0}=0.2$ (continuous line), $r_{0}=0.3$ (doted line) and
$r_{0}=0.4$ (dashed line). \newline
\textbf{Right diagram:} $q=1$, $\protect\alpha =0.001$ (continuous line), $%
\protect\alpha =0.005$ (doted line) and $\protect\alpha =0.009$ (dashed
line).}
\label{Fig8}
\end{figure}

In this case, the divergency of the deficit angle is located at
\begin{equation}
\left. r_{0}\right\vert _{\delta \varphi \longrightarrow \infty }=\frac{%
\left( 2\mathcal{X}\right) ^{\frac{1}{3}}+2q+2q^{2}\left( 4\mathcal{X}%
^{-1}\right) ^{\frac{1}{3}}}{3\Lambda l^{2}}q,
\end{equation}%
\begin{equation}
\mathcal{X}=q\left[ 4q-27\Lambda ^{2}l^{2}\alpha +3\Lambda l\sqrt{3\alpha
\left( 27\Lambda ^{2}l^{2}\alpha -8q^{2}\right) }\right] ,
\end{equation}%
and the root of the deficit angle is located at
\begin{equation}
\left. r_{0}\right\vert _{\delta \varphi =0}=\frac{\left( 9\sqrt{2}\mathcal{Y%
}\right) ^{\frac{2}{3}}q+2q^{2}\left( 3\mathcal{Y}^{\frac{1}{3}}+6^{\frac{2}{%
3}}q\right) }{9l(\Lambda l+2)\mathcal{Y}^{\frac{1}{3}}},
\end{equation}%
where
\begin{equation}
\mathcal{Y}=q\left[ \frac{4}{3}q^{2}-9\alpha \left( \Lambda l+2\right)
^{2}+9\left( \Lambda l+2\right) \sqrt{\alpha \left[ \alpha \left( \Lambda
l+2\right) ^{2}-\frac{8q^{2}}{27}\right] }\right] .
\end{equation}

Considering the importance of deficit angle and its contribution to geometry
of solutions, we plot various graphs (Fig. (\ref{Fig8})) for studying the
effects of variation of parameters on deficit angle.

For the case of additional correction as nonlinear electromagnetic field,
following effects were seen. As for the case of variation of charge (Fig. %
\ref{Fig8} left), deficit angle has a singularity. In other words, there is
a divergency in which before and after that deficit angle is showing
different behavior. Before divergency deficit angle is a decreasing function
of nonlinear parameter and there is a root for it and a region in which it
has negative value. Whereas after divergency the deficit angle is a
decreasing function of nonlinearity and it is always positive which is
located in the region of non acceptable values. In other words, the values
of the deficit angle after divergency are not in the upper bound limit of
the deficit angle. The place of this divergency is a decreasing function of
charge.

Next, as for the effects of $r_{0}$ Fig. \ref{Fig8} (middle) is plotted in
which, there is a singularity for deficit angle. The place of divergency is
an increasing function of the $r_{0}$. This behavior is opposite to the
behavior of deficit angle for variation of electric charge.

The effects of variation of nonlinear parameter is seen in Fig. \ref{Fig8}
(right). Plotted graph for deficit angle versus $r_{0}$ shows that in
essential, there is a divergency for calculated values of deficit angle. The
place of this divergency is an increasing function of nonlinear parameter
which means that, as nonlinear parameter increases, the place of this
divergency will move to higher values of $r_{0}$. In case of deficit angle
versus $r_{0}$, the behavior of system is quite different. Opposite to the
previous case, in this plot before singularity, deficit angle is an
increasing function of $r_{0}$ which is higher that upper bound limit for
deficit angle and after that, there will be a region of negative deficit
angle. This region and its related root are increasing functions of
nonlinearity parameter. The lowest value of deficit angle belongs to the
highest value of nonlinear parameter.

\subsubsection{Spinning AC Magnetic Solution}

In order to add angular momentum to the spacetime, we insert Eq. (\ref%
{rotating boost}) into Eq. (\ref{change coordinate metric}) and we obtain
the rotating metric (\ref{rotating metric}), where $g(r)$ is the same as $%
g(r)$ which is given in Eq. (\ref{change Lagrangian metric}). The
electromagnetic filed components become
\begin{equation}
F_{rt}=-\frac{a}{\Xi l^{2}}F_{r\varphi }=\frac{a}{\Xi l^{2}}\left( \frac{q}{%
\left( r^{2}+r_{0}^{2}\right) ^{1/2}}-\frac{4q^{3}\alpha }{\left(
r^{2}+r_{0}^{2}\right) ^{3/2}l^{2}} \right)+O\left( \alpha ^{2}\right) .
\end{equation}

The electric charge can be found by calculating the flux of the electric
field at infinity, yielding
\begin{equation}
Q=\frac{q}{2\pi }\sqrt{\Xi ^{2}-1}.
\end{equation}

On may note that the electric charge is proportional to the rotation
parameter and for the static case ($\Xi =1$) is zero. Also, one can show
that the mass and the angular momentum of the solution are same as those in
Eqs. (\ref{MPMI}) and (\ref{JPMI}), respectively.

At last, in order to obtain an insight regarding the negative deficit angle
to geometrical structure of the solutions, we first give a description
regarding positive deficit angle. The conic-like structure of the solutions
is due to absence of specific value of angle that was cut from spacetime.
This angle is deficit angle and has a positive value. In case of negative
value of deficit angle, it is like an added segment to the spacetime. This
adding will change the structure of the spacetime to a saddle-like cone (for
more details see Fig. $2$ in Ref. \cite{LR}). This negative deficit angle is
known as surplus angle. It is worthwhile to mention that there is an upper
bound for positive deficit angle whereas such bound does not exist for
negative values of deficit angle.

\section{Conclusions}

In this paper, we considered different nonlinear theories of electrodynamics
and study their three dimensional magnetic solutions. Although these
theories of nonlinearity are generalization of Maxwell theory, in essence
they are describing different phenomena. The obtained metric functions for
these nonlinear theories showed quite different structures for magnetic
solutions which in result enforcing their own conditions.

The primitive motivation of considering the mentioned metric was obtaining
magnetic solutions through topological defects. In other words, the obtained
values are representing topological defects. This conclusion is valid
because of the geometrical structure of obtained solutions and the important
property known as deficit angle. The deficit angle shows that the object
that we are studying is not usual geometrical object. In case of obtained
solutions in this paper, due to being three dimensional, their $t=cte$ and $%
r=cte$ geometry is a ring-like. Its shape and general properties such as
area are described and determined by the value of deficit angle.

At first we focused on the energy density. Studying energy conditions in
context of this spacetime, revealed the fact that PMI, LNED and ENED
theories satisfy null, weak, strong and dominant energy conditions. In case
of PMI theory, dominant energy conditions put a restriction on valid range
of $s$ parameter ($\frac{1}{2}<s\leq 1$). No restriction was observed for
LNED and ENED. Next we studied the effects of different nonlinear theories
on energy density and compare them with Maxwell theory. Interestingly for
case of PMI theory, we had two sets of behavior. In general the energy
density was an increasing function of $s$. Considering the fact that $s=1$
is denoted as Maxwell theory, we \ found that in case of $\frac{1}{2}<s<1$ ($%
s>1$) the concentration volume of energy density was smaller (larger)
comparing to Maxwell theory. On the other hand, BI-types theories (LNED and
ENED) had larger energy density than Maxwell theory. In general in these two
theories energy density was a decreasing function of nonlinearity parameter.
Therefore, considering the fact that for large values of nonlinearity
parameter, these two theories reduce to Maxwell theory, one expect that the
lowest energy density between these theories belongs to Maxwell theory,
which is consistent with obtained results.

Here we are encountering another important difference between PMI theory and
BI-types theories. In essence, the generalization of Maxwell theory to
nonlinear theories of BI-types causes an increase in energy density. This
increase indicates that the distribution of matter filed in these nonlinear
theories is more concentrated comparing to Maxwell theory. On the other
hand, for PMI theory two scenarios is possible. In one ($\frac{1}{2}<s<1$),
considering this nonlinear theory causes to decrease of energy density. In
other words, in this case the distribution of matter filed is less
concentrated comparing to Maxwell theory. On the other hand, for the one ($%
s>1$), the energy density becomes larger comparing to Maxwell theory. This
two different behavior is a unique characteristic of PMI theory and
emphasizes the different nature of this theory from BI-types. If one
consider dominant energy condition and its restrictions on theories of
nonlinearity as dominant limitations, PMI theory only increases the energy
density whereas the BI-types increases the energy density. In the other
words, these two classes of nonlinear theories have opposite effects on
energy density.

In essence, PMI theory is a different theory comparing to other ones in
conditions and evaluated values. The existence of $s$ as a power makes the
magnetic solution related to it more sensitive to variation of $s$ comparing
to variation of nonlinearity parameter in other theories. The places of
deficit angel root and divergency were highly sensitive to variation of $s$.
Due to structure of this theory two behaviors were seen for different values
of $s$ which is a characteristic that only belongs to this theory. These
different behaviors add another free parameter to this theory and make it
possible to consider two approaches for studying magnetic solutions. It also
states that in considering this theory, one must take this undeniable
important property into consideration for studying solutions and their
properties such as conserved quantities and their general behaviors.

In addition, this fact is of importance to mention that in usual charged
three dimensional solutions, one expects the rise of logarithmic function of
radial coordinate in metric function. This function was seen in BI-type
nonlinear theories whereas for the case of PMI, interestingly, only for
certain value of $s$ this function was seen. This fact emphasis another
fundamental difference between this theory of nonlinear electromagnetic
field and BI-type ones.

As for the AC theory, due to consideration of nonlinear parameter as a
correction to Maxwell theory, there was a restriction of considering only
small values of nonlinearity parameter. Interestingly, in this theory, the
existence of divergency was seen for deficit angle.

Remarkably, for case of LNED no singular point, hence no divergency was
seen. Contrary to AC theory, this theory presented smooth and divergence
free behavior for deficit angle. The obtained values of deficit angle for
this theory were real and the only restriction that one may confront comes
from the logarithmic part of solution which in plotted graphs for deficit
angle no effect of this restriction was seen. Although both LNED and AC
theories are in essence BI-types, this behavior is showing an important fact
that they are in case of topological defects and magnetic solutions are
describing completely different phenomena and they are independent of each
other. The same property was seen for the case of exponential form.

It is notable to mention the fact that in plotted graphs of Maxwell, no
singularity was seen. In fact, calculated values of divergence point showed
that there are two divergence points that in AdS spacetime they are not
real. In other words, in case of AdS spacetime, deficit angle is divergence
free. Opposite to the case of divergency, we found a relation for roots in
this case which indicated three different possible cases: two roots, one
extreme root and no root.

One of important issue that must be taken into consideration is the
existence of roots for deficit angle. The existence of root for deficit
angle states that no contributing to structure of magnetic solution exists.
In other words, the object that we are studying in these special cases are
not cosmological (topological) defects and they do not have the property of
being cosmological defects. If we consider the cosmological defects as
dynamic objects that their parameters may vary through time, one may say
that for special values of parameters, the object will change into another
astrophysical object (no deficit angle is seen). But this idea is only
acceptable if the root of deficit angle is extreme or the region in which
deficit angle is negative.

Also, the existence of negative values of deficit angle poses another
important issue. The structure of magnetic solution and the meaning of
having negative deficit angle is something that must be taken into
consideration and studied in more details.

One may interpret that roots of deficit angle may present the phase
transition for these astrophysical objects and the negative values of
deficit angle are representing another phase for them. Or one may say that
negative and positive values of deficit angle are representing two different
types of defects. The roots are places where these phase transitions take
place. Considering the fact that in calculation of deficit angle, one is
using second order derivation of metric function with respect to radial
coordinate (see for example chapter $9$ of Ref. \cite{dInverno}) and if one
consider the metric function as a potential, it is arguable that the roots
of deficit angle are representing phase transition. On the other hand,
considering the concept of divergency of potential as a point of phase
transition, one may argue that existence of divergency in deficit angle is
representing phase transition. Therefore, one may state that instead of
taking roots of deficit angle as phase transition points, singular points
must be taken into consideration as phase transition points. These phase
transitions may be geometrical types of transitions. In other words, the
shape of the object may only change, not its physical being change into
another thing. But this idea is debatable if one consider roots of deficit
angle as phase transition. It is due to fact that topological property which
describe the shape of the magnetic object will be quite different before and
after phase transition in which the sign of deficit angle will change. In
some of the nonlinear theories and Maxwell one no singularity was seen which
state that in concept of considering divergency as a phase transition, these
theories are in fact without phase transition. But as it was mentioned
before, in case of Maxwell theory, the background spacetime (AdS/dS) plays
the crucial role. In AdS spacetime there is no divergency and for dS
spacetime one can find divergence points and therefore it may have phase
transition. But if one consider roots as phase transition in both spacetime,
phase transitions take place.

Another interesting issue comes from studying Fig. (\ref{Fig33}). In the
absence of charge, $q=0$, the deficit angle could be non zero. By adding
charge to solutions and increasing it, the deficit angle increases and
general behavior of it is also modified. This shows the fact that
contribution of charge to deficit angle is of an increasing factor. In other
words, electromagnetic field will increase the value of deficit angle.

Finally it is quite important to mention the fact that the only non zero
component of considered gauge potential in case of these topological defects
were spatial one which was considered as provider of magnetic field. By
applying the mentioned transformation and changing metric from static to
rotating one, another component was added to electromagnetic field tensor
which was the well-known provider of electric field. Obtained values for
this electric part of electromagnetic field tensor were functions of
rotating parameter and in case of setting rotating parameter equal to zero,
these electric field would vanish. This fact shows that obtained values are
essentially magnetic solutions.

\begin{acknowledgements}
We thank Shiraz University Research Council. This work has been
supported financially by Research Institute for Astronomy and
Astrophysics of Maragha.
\end{acknowledgements}


\begin{thebibliography}{999}
\bibitem{DW} Two dimensional membranes that form when a discrete symmetry is
broken at a phase transition. These walls resemble the walls of a
closed-cell foam, dividing the universe into discrete cells.
\bibitem{Mon} Cube-like defects that form when a spherical symmetry is
broken and are predicted to have magnetic charge, either north or south. The
most well-known monopole is magnetic monopole.
\bibitem{Tex} Form when larger, more complicated symmetry groups are
completely broken. They are not as localized as the other defects and are
unstable. Other more complex hybrids of these defect types are also possible.
\end{thebibliography}
\end{document}